\documentclass[floats,aps,amsmath,amssymb]{revtex4}
\usepackage{graphicx}
\usepackage{subfigure}
\usepackage{hyperref} 
\usepackage{bm}
\usepackage{float}
\usepackage{yfonts}
\usepackage{xcolor}
\usepackage{dsfont}
 \usepackage{ulem}

\hypersetup{
    colorlinks,
    linkcolor={blue!80!black},
    citecolor={blue!80!black},
    urlcolor={blue!80!black}
}
\usepackage[mathcal]{eucal}

\begin{document}

\title{Transient dynamics in interacting nanojunctions within self-consistent perturbation theory}
\author{R. Seoane Souto$^1$, R. Avriller$^2$, A. Levy Yeyati$^1$ and A. Mart\'in-Rodero$^1$}
\affiliation{$^1$Departamento de F\'{i}sica Te\'{o}rica de la Materia Condensada,\\
Condensed Matter Physics Center (IFIMAC) and Instituto Nicol\'{a}s Cabrera,
Universidad Aut\'{o}noma de Madrid E-28049 Madrid, Spain}
\affiliation{$^2$Univ. Bordeaux, CNRS, LOMA, UMR 5798, F-33405 Talence, France}
\date{\today}

\begin{abstract}
We present an analysis of the transient electronic and transport properties of a nanojunction in the presence of electron-electron and electron-phonon interactions. 
We introduce a novel numerical approach which allows for an efficient evaluation of the non-equilibrium Green functions in the time domain. Within this approach we implement different self-consistent 
diagrammatic approximations in order to analyze the system evolution after a sudden connection to the leads and its convergence to the steady state.
 These approximations are tested by comparison with available numerically exact results, showing good agreement even for the case of large
interaction strength. In addition to its methodological advantages, this approach allows us to study several issues of broad current interest like the build up in time 
of Kondo correlations and the presence or absence of bistability
associated with electron-phonon interactions. We find that, in general, correlation effects tend to remove the possible appearance of charge bistability.
\end{abstract}

\maketitle

\section{Introduction}
For decades, studies of quantum transport in nanoscale devices have mainly focused on 
steady state properties \cite{nazarov2009quantum}. While the potential interest of transient dynamics 
was pointed out long ago \cite{Blandin_JP1976,Jauho_PRB1994} such studies 
have recently received an increasing attention in connection with
advances in experimental techniques for time-resolved measurements \cite{Feve_Science2007,Vink_APL2007,Flindt_PNAS2009,Terada_NatPhot2010,Loth_Science2010,Latta_Nature2011,
Yoshida_NatNano2014,Otsuka_SCR2017,Karnetzky_axiv,Du_arXiv2018}. These 
studies are also motivated by the important technological goal of increasing the operation speed of devices while reducing their energy consumption. 
Moreover, studies of the transient dynamics after a quench of a given parameter are currently undertaken in many fields of physics ranging from cold atoms 
\cite{Hung_science2013,Schreiber_science2015}, correlated materials \cite{Cazadilla_PRL2006}, dynamical phase transitions \cite{Heyl_PRL2013} and, more generally, in connection to the question on the 
existence of a well defined stationary state for any 
given model of interacting particles \cite{Eisert_NatPhy2015}.\\

On the theoretical side transport transient dynamics has been addressed using different methods valid for different regimes. Thus, the scattering approach or the non-equilibrium Green 
function formalism have been used for describing the dynamics in the coherent non-interacting regime \cite{Cini_PRB1980,Stefanucci_PRB2004,Albert_PRL2011,Albert_PRL2012,Dasenbrook_PRL2014,Tang_PRB2014,
Tang_PRB2014_2,Ridley_PRB2015,Murakami_PRB2015,Souto_Fort2017,Odashima_PRB2017,Covito_arXiv2018}. However, the inclusion of interactions is essential to analyze the transport dynamics 
through localized states, as is the case of molecular junctions or semiconducting quantum dots. 
For these cases, rate equations approaches, adequate for a sequential tunneling regime, have been extensively used \cite{Kambly_PRB2011,Stegmann_PSS2017}. The most interesting and general 
coherent-interacting regime constitutes a great theoretical challenge. This regime has been addressed
using several complementary approaches: diagrammatic techniques \cite{Meir-Wingreen_PRL1999,Plihal_PRB2005,Goker_JPCM2007,Schmidt_PRB2008,Komnik_PRB2009,Myohanen_PRB2009,Myohanen_PRB2012,Perfetto_PRB2013,
Riwar_Schmidt,Latini_PRB2014,Vinkler_PRB2014,Souto_PRB2015,Chen_PRB2016,Tang_NJP2017,Tang_PRB2017},  quantum Monte-Carlo 
\cite{Muhlbacher_PRL2008,Albrecht_PRB2012,Cohen_PRB2013,Cohen_PRL2014,Hartle_PRB2015,Klatt_PRB2015,Albrecht_PRB2015,Ridley_PRB2018}, 
time-dependent NRG \cite{Anders_PRB2005,Anders_PRB2006,Heidrich_PRB2009,Eckel_NJP2010,Edelstein_PRB2012,Guttge_PRB2013,Nghiem_PRB2014,Nghiem_PRL2017},
time-dependent DFT \cite{Zheng_PRB2007,Kurth_PRL2010,Uimonen_PRB2011,Khosravi_PRB2012,Kwok_FPh2014,Dittman_PRL2018,Kurth_EPL2018} 
among others \cite{Wang_JCP2003,Weiss_PRB2008,Kennes_PRB2012,Kennes_PRB2012_2,Perfetto_JCE2015}.
However, all of these techniques as they are actually implemented have some limitations. For instance, numerically exact methods like quantum Monte-Carlo are 
strongly time-consuming, require finite temperature and typically do not allow to reach long time scales.  Similar concerns can be applied to the case of 
time-dependent NRG. \\

This situation suggests the convenience of revisiting perturbative diagrammatic
methods for analyzing transport transient dynamics in interacting nano-scale devices. 
Although these methods have been partially explored in previous works \cite{Riwar_Schmidt,Schmidt_PRB2008}, these implementations did not, in general, include self-consistency which can become of essence 
in order to increase the accuracy and range of validity of these methods. Moreover, in the case of models including electron-phonon interactions further methodological developments are needed in order to 
take into account properly the dynamical build up of a non-equilibrium phonon distribution.\\

In this work we present an efficient algorithm for the integration of the 
time-dependent Dyson equation for the non-equilibrium Green functions 
applied to different models of correlated nano-scale systems, including
electron-electron and electron-phonon interactions. To deal with 
these correlations we use a diagrammatic expansion of the system
self-energies at different levels of approximation including self-consistency effects. In the case of electron-phonon interactions we introduce novel theoretical tools for solving the Dyson 
equations associated with the phonon propagator in order to account properly for the build up of a non-equilibrium phonon population.
As a check of these approximations we study the convergence of the system properties like mean
charge, current and spectral density to their stationary values and also compare
them to available numerically exact results. When not available we have implemented our own NRG calculations. 
We show how this time-dependent approach is quite convenient for including self-consistency in a 
straightforward way. We exemplify the use of this methodology to investigate the 
issue of bistability for the molecular junction, demonstrating how the inclusion of 
correlation effects beyond the mean-field approximation tends to eliminate the bistable behavior of
 charge and current for certain parameter regimes.\\    

The paper is organized as follows: In Sec. \ref{sec::formalism} we introduce the formalism and the numerical techniques used for computing the transient 
electronic and transport properties; in Sec. 
\ref{sec::U} we analyze the dynamics of a system with strong electron-electron
interactions taking the non-equilibrium Anderson model as a paradigmatic example.
Sec. \ref{sec::eph} is devoted to the study of the transient properties in
the presence of electron-phonon interactions by means of 
the spinless Anderson-Holstein model. 
In Sec. \ref{sec::AH} we consider a situation where both electron-electron
and electron-phonon interactions are present using the spin-degenerate 
Anderson-Holstein model.
Finally we present the conclusions and provide a brief overlook of
our main results in Sec. \ref{sec::conclusions}.

\section{Keldysh formalism for the transient regime}
\label{sec::formalism}
For describing a nanoscale central region coupled to metallic electrodes we consider a model Hamiltonian of the form
$ \hat{H}=\hat{H}_{leads}+\hat{H}_c+\hat{H}_{T}+\hat{H}_{int}$, where 

\begin{eqnarray}
\hat{H}_{leads}=\sum_{k\sigma,\nu}\epsilon_{k\sigma,\nu}c^{\dagger}_{k\sigma,\nu}c_{k\sigma,\nu},\quad
\hat{H}_{c} = \sum_{\sigma}\epsilon_0c^{\dagger}_{0\sigma} c_{0\sigma}, \quad
\hat{H}_T=\sum_{k\sigma,\nu}\left[v_{k\sigma,\nu}(t)c^{\dagger}_{k\sigma,\nu}c_{0\sigma}+\mbox{h.c}\right], 
\end{eqnarray}
where $c_{k\sigma,\nu}$, with $\nu=L\,(R)$ labeling the left (right) electrode, and $c_{0\sigma}$ are annihilation operators for electrons in the leads and in the central region respectively and
 $v_{k\sigma,\nu}(t)$ is the tunneling amplitude which will depend on time.
The two electrodes can be kept at different chemical potentials via a constant bias voltage $eV=\mu_L-\mu_R$. For simplicity the central region will consist of a single 
quantum level denoted by $\epsilon_0$. 
The last term, $\hat{H}_{int}$, in $\hat{H}$ describes the many body interactions in the central region, which we shall specify later. Hereafter we assume $e=\hbar=k_B=1$.\\

In what follows we will consider the wide-band approximation for the electrodes. Within this approximation the tunneling rates can be taken as constants, 
$\Gamma_\nu=\pi\sum_k|v_{k\sigma,\nu}|^2\delta(w-\epsilon_{k\sigma,\nu})\sim\pi|v|^2\rho_F$, where $\rho_F$ is the density of states at the Fermi edge, the resonant level width being 
$\Gamma=\Gamma_L+\Gamma_R$. Our aim is to analyze the transient dynamics of such a correlated system after a sudden quench of the coupling to the electrodes at an initial time that we take at $t=0$. 
Thus, $v_{k\sigma,\nu}(t)=\theta(t)v_{k\sigma,\nu}$, which allows us to define a time-dependent tunneling rate $\Gamma(t)=\theta(t)\Gamma$. Although this work is focused on this sudden connection 
case, more general time-dependent Hamiltonians could be considered within the formalism presented below.\\

\begin{figure}
   \begin{minipage}{0.48\textwidth}
     \centering
     \includegraphics[width=.9\linewidth]{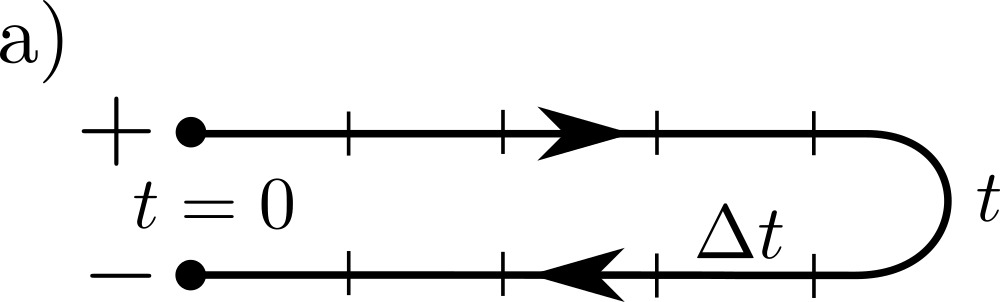}
   \end{minipage}\hfill
   \begin {minipage}{0.48\textwidth}
     \centering
     \includegraphics[width=.9\linewidth]{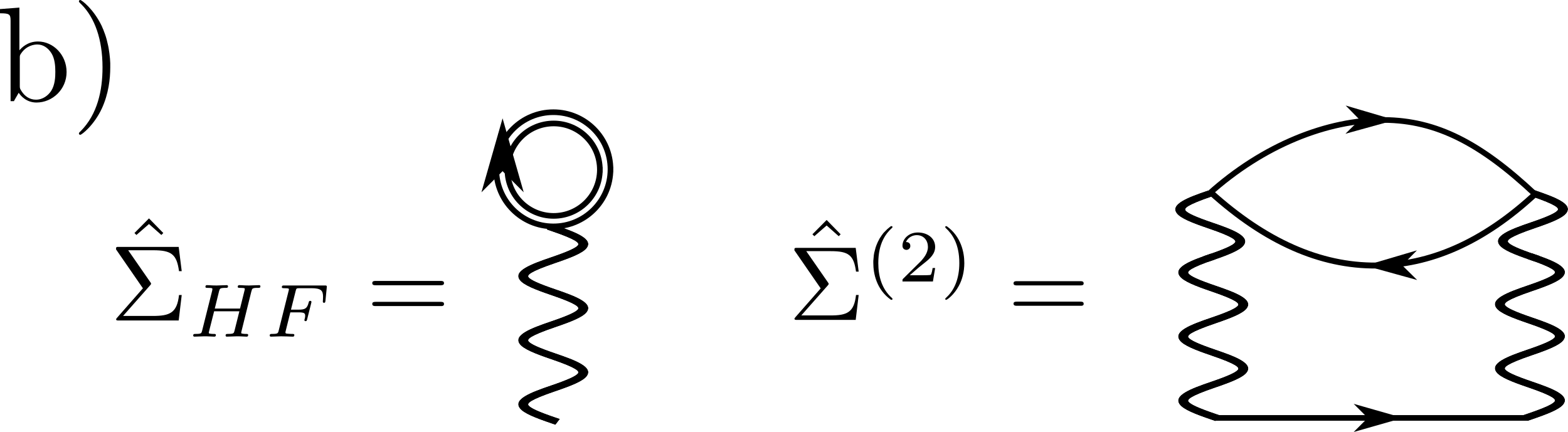}
   \end{minipage}
\caption{a): Keldysh contour considered to analyze the transient regime, $\Delta t$ being the time step in the discretized calculation of the time dependent Green function. b): Self-energy
diagrams for the Coulomb interaction up to second order. The solid line represents the electron propagator and the wavy line the interaction. In the HF diagram the double line indicates the charge 
calculated using the dressed propagators.}
    \label{contour_Diagrams_U}
\end{figure}

The dynamical electronic and transport properties can be obtained from the central level Green functions in Keldysh space, 
$\hat{G}_\sigma(t,t') = -i \langle \hat{T}_c c_{0\sigma}(t) c^{\dagger}_{0\sigma}(t') \rangle$, where $\hat{T}_c$ is the 
chronological time-ordering operator along the Keldysh contour \cite{Keldysh_JETP1965} (see Fig. \ref{contour_Diagrams_U} a). In the absence of interactions
the problem is exactly solvable even in the presence of an arbitrary time dependent
potential \cite{Blandin_JP1976,Jauho_PRB1994}.
However, in the presence of interactions the problem of obtaining the dynamical behavior of the system usually becomes extraordinarily demanding. On the one side, there is the 
problem of finding an appropriate 
treatment of correlation effects by means of an adequate self-energy. This is not always a simple task in the dynamical problem. On the other hand, even if an appropriate 
self-energy is found, the 
numerical solution of the Dyson equation for the Keldysh propagators (which in the time domain becomes an integral equation) is a formidable numerical problem.\\

In this section we present an efficient numerical procedure for the calculation of the Keldysh propagators in the transient regime. It allows us to obtain accurate results for the electronic and 
transport properties such as the central region charge and current. The power of the method is additionally checked by analyzing the convergence of these quantities (together with the central region
spectral density) to their expected stationary values.\\

We start from the Dyson equation for the central level Green function in Keldysh space, which can be formally inverted
\begin{equation}
 \hat{G}_\sigma=\left[\hat{g}^{-1}_\sigma-\hat{\Sigma}_{\sigma,T}-\hat{\Sigma}_{\sigma,int}\right]^{-1},
 \label{Dyson}
\end{equation}
where $\hat{g}^{-1}_{\sigma}$ is the inverse free electron propagator of the uncoupled central level, $\hat{\Sigma}_{\sigma,T}$ the tunneling self-energy and $\hat{\Sigma}_{\sigma,int}$ the 
interaction self-energy. Interactions mixing the spin degree of freedom could be also included in the equation as discussed in Refs. \cite{Souto_PRL2016,Souto_PRB2017}.
Eq. (\ref{Dyson}) can be numerically solved by discretizing time in the Keldysh contour (see Fig. \ref{contour_Diagrams_U} a). From now on the discretized matrix propagators and 
self-energies will be denoted in boldtype. The inverse free level Green function discretized on the contour is then given by \cite{kamenev_book}
\begin{equation}
i {\bold{g}}^{-1}_\sigma = \left(\begin{array}{cccc|cccc} -1 & & & & & & & -\rho_\sigma \\
h_- & -1 & & & & & &  \\
& h_- & -1 & & & & &  \\
& & \ddots & \ddots & & & &  \\
\hline  
&  & & 1 & -1 & & &  \\
&  & & & h_+ & -1 & &   \\
&  &  &  & & \ddots & \ddots &  \\
&  &  & & & & h_+ & -1 \end{array} \right)_{2N\times2N} \; .
\label{kamenev}
\end{equation}
where $h_{\pm} = 1 \mp i\epsilon_0 \Delta t$, $\Delta t$ indicates the time step in the discretization with $N=t/\Delta t$. In this expression the initial level charge is determined by 
$n_\sigma(0) = \rho_\sigma/(1+\rho_\sigma)$. Note that the discretization over the contour is made starting from $t=0$ to the final time through the positive Keldysh branch and returning to $t=0$ through 
the negative one.\\


The time-dependent tunneling self-energies can be evaluated straightforwardly and at zero temperature have the simple form \cite{Souto_PRB2015}
\begin{eqnarray}
 \Sigma^{+-}_{T,\sigma}(t,t')=\frac{\Gamma}{\pi}\sum_\nu\frac{e^{-i\mu_\nu(t-t')}-e^{iD(t-t')}}{(t-t')},\qquad
 \Sigma^{-+}_{T,\sigma}(t,t')=\frac{\Gamma}{\pi}\sum_\nu\frac{e^{-i\mu_\nu(t-t')}-e^{-iD(t-t')}}{(t-t')},
\end{eqnarray}
$2D$ being the leads bandwidth. Alternatively, it is possible to take the limit
$D \rightarrow \infty$ provided that
a finite temperature, taken as
the smallest energy parameter, is introduced (see Ref. \cite{Souto_PRB2015}).
In all the results given below we consider this infinite bandwidth limit except 
when comparing with numerically exact methods where an energy cutoff 
with a precise value is used.  
The other Keldysh self-energy components are then given by
\begin{eqnarray}
 \Sigma_{T,\sigma}^{++}(t,t')=-\theta(t-t')\Sigma_{T,\sigma}^{-+}(t,t')-\theta(t'-t)\Sigma_{T,\sigma}^{+-}(t,t'),\qquad
 \Sigma_{T,\sigma}^{--}(t,t')=-\theta(t-t')\Sigma_{T,\sigma}^{+-}(t,t')-\theta(t'-t)\Sigma_{T,\sigma}^{-+}(t,t')\,,
\label{Sigma_Keldysh}
\end{eqnarray}
where $\theta(t)$ is the Heaviside step function. Notice that there is an ambiguity in the definition of these self-energies at equal times. It turns out that the different 
possible choices in the 
definition of $\Sigma_{T,\sigma}^{++}(t,t)$ and $\Sigma_{T,\sigma}^{--}(t,t)$ can significantly affect the convergence and stability of the system properties with time. Although the precise value 
of $\Sigma_{T,\sigma}^{++}(t,t)$ and $\Sigma_{T,\sigma}^{--}(t,t)$ depends on the whole energy range of the leads density of states, if one is not interested in the dynamics on 
time scales smaller than $1/D$ there is freedom to choose this value. We have found that the most stable algorithm corresponds to the choice
\begin{equation}
 \Sigma_{T,\sigma}^{++}(t,t)=\Sigma_{T,\sigma}^{--}(t,t)=-\frac{\Sigma_{T,\sigma}^{+-}(t,t)+\Sigma_{T,\sigma}^{-+}(t,t)}{2}.
\label{half_criteria}
\end{equation}
We have checked that this choice appropriately recovers the correct stationary limit and perfectly reproduces the transient behavior in the cases where an analytic expression is available 
(see section \ref{subsec::U_HF}).\\

The evaluation of the interaction self-energy will be discussed in sections \ref{sec::U}-\ref{sec::AH} for the cases of electron-electron and electron-phonon interactions. For computing the correlation 
part of the interaction self-energy we also find that the most stable algorithm consists on the calculation of the non-diagonal Keldysh components
($\hat{\Sigma}_{int}^{+-}(t,t')$ and $\hat{\Sigma}_{int}^{-+}(t,t')$) and then using the relations of Eqs. 
(\ref{Sigma_Keldysh},\ref{half_criteria}) for the diagonal ones.\\

The self-energies are then evaluated in the discrete time mesh (left Fig. \ref{contour_Diagrams_U} a). The propagators in Keldysh space can now be obtained by numerically inverting the matrix
\begin{equation}
 \hat{\bold{G}}^{-1}_\sigma=\hat{\bold{g}}^{-1}_\sigma-(\Delta t)^2\left(\hat{\bold{\Sigma}}_{T,\sigma}+\hat{\bold{\Sigma}}_{int,\sigma}\right).
 \label{Dyson_discretized}
\end{equation}
Notice the factor $(\Delta t)^2$ introduced by the discretization procedure.\\

The knowledge of 
$\hat{G}_\sigma(t,t')$ enable us to calculate the evolution with time of the electronic and transport properties of the system such as the central level charge, the spectral density and the current. Thus,
the level charge can be calculated as $n_\sigma(t)=iG_{\sigma}^{+-}(t,t)$, while the current through the interface between the central region an the electrodes is given by
\begin{eqnarray}
 I_\nu=\sum_\sigma\int_{0}^{t}\left[G^{+-}_\sigma(t,t_1)\Sigma^{-+}_{T,\sigma\nu}(t_1,t)-G^{-+}_\sigma(t,t_1)\Sigma^{+-}_{T,\sigma\nu}(t_1,t)\right]dt_1\;.
\end{eqnarray}

Finally, following Refs. \cite{Albrecht_PRB2015,CohenMillis_PRB2016}, it is possible to define a time dependent auxiliary spectral density function per spin $A_\sigma(\omega,t)$ by calculating the current
to weakly coupled probes and which tends to the correct stationary value at large times $A_\sigma(\omega)=\mbox{Im}\left[G^{A}_\sigma(\omega)-G^{R}_\sigma(\omega)\right]/2\pi$. For the present system we 
have
\begin{equation}
  A_\sigma(\omega,t)=\mbox{Im}\int_{0}^{t}dt'\frac{e^{-i\omega\, (t-t')}}{2\pi}\left[G^{+-}_\sigma(t',t)-G^{-+}_\sigma(t',t)\right],
\end{equation}
and the spin averaged spectral density as $A(\omega,t)=\sum_\sigma A_\sigma(\omega,t)/2$.

\section{Electron-electron interaction: the Anderson model}
\label{sec::U}
In this section
we will consider the Anderson model \cite{Anderson_PR1961} consisting of a single spin degenerate level with on-site electron-electron repulsion, coupled to metallic electrodes. 
The interaction term in the Hamiltonian of Sec. \ref{sec::formalism} is given by $\hat{H}_{e-e}=U\hat{n}_\uparrow \hat{n}_\downarrow$, where $\hat{n}_\sigma=c^{\dagger}_{0\sigma} c_{0\sigma}$ and $U$ is 
the local Coulomb repulsion.

\subsection{Hartree-Fock solution}
\label{subsec::U_HF}
The dynamical Hartree-Fock (HF) solution of the Anderson model provides an ideal test of the accuracy of the numerical method presented in sect. \ref{sec::formalism} as in this case the time-dependent problem can be exactly solved \cite{Jauho_PRB1994,Blandin_JP1976}.
Thus, within this approximation, the model becomes an effective single electron problem with a spin and time-dependent central level
\begin{equation}
 \epsilon_\sigma(t)=\epsilon_0+Un_{\bar{\sigma}}(t)\;,
\end{equation}
where $n_\sigma(t)$ is the central level occupation per spin. As commented in the
previous section, the problem of an impurity level in a time-dependent potential coupled to metallic leads is exactly solvable using the Keldysh
method. For the HF case addressed in this paper, the Keldysh Green function can be written in a very compact way as
 \begin{equation}
 G_{HF,\sigma}^{+-}(t,t')=\theta(t)\theta(t')ie^{-i\left[t\bar{\epsilon}(t)-t'\bar{\epsilon}(t')\right]}e^{-\Gamma(t+t')}\left\{n_\sigma(0)+\frac{1}{\pi}\int{ d\omega \left[\sum_{\nu=L,R}\Gamma_\nu f_\nu(\omega)\right]}g_\sigma(\omega,t)g^{*}_\sigma(\omega,t')\right\}\;,
\label{G_HF}
\end{equation}
where
\begin{eqnarray}
 \bar{\epsilon}_\sigma(t)=\frac{1}{t}\int_{0}^{t} d\tau \epsilon_\sigma(\tau)\,,\qquad
 g_\sigma(\omega,t)=\int_{0}^{t} d\tau e^{-i\left[\omega+i\Gamma-\bar{\epsilon}_\sigma(\tau)\right]\tau}\;.
\end{eqnarray}
The time evolution of the central level occupation is then obtained as $n_\sigma(t)=iG_{\sigma,HF}^{+-}(t,t)$ and has the form
\begin{eqnarray}
 n_\sigma(t)=e^{-2\Gamma t}\left\{n_\sigma(0)+\int{\frac{d\omega}{\pi} \left[\sum_{\nu=L,R}\Gamma_\nu f_\nu(\omega)\right]}{\left|g_\sigma(\omega,t)\right|^2}\right\}.
\label{charge_HF}
\end{eqnarray}

One can compare the result of the numerical method proposed in Sec. \ref{sec::formalism} with Eq. (\ref{charge_HF}). In the HF approximation the self-energy is given by the left diagram of Fig. 
\ref{contour_Diagrams_U} b) and has the form
\begin{equation}
 \Sigma_{HF,\sigma}^{\alpha\beta}(t,t')=\alpha U\,n_{\bar{\sigma}}(t)\delta(t-t')\delta_{\alpha\beta}\;,
\label{Anderson_HF}
\end{equation}
where $\alpha,\beta=\pm$ are the Keldysh branch indexes. Notice that the Dirac delta in the previous equation is converted to a Kronecker $\delta$ function, including an additional $1/\Delta t$ factor 
when discretizing in the time mesh.
We can now obtain the propagators in the HF approximation by inverting
\begin{equation}
\hat{\bold{G}}_{HF,\sigma}^{-1}=\hat{\bold{g}}^{-1}_\sigma-(\Delta t)^2\left(\hat{\bold{\Sigma}}_{T,\sigma}+\hat{\bold{\Sigma}}_{HF,\sigma}\right),
\label{G_HF_U}
\end{equation}
and following the numerical procedure presented in the previous section. The dynamical properties of the system can be now calculated from $\hat{G}_{HF,\sigma}$.\\

It is worth remarking that the self-consistency condition on the charge in this approximation is particularly straightforward as it is simply achieved by storing the charge values obtained in the discrete 
mesh by inverting Eq. (\ref{G_HF_U}) at each time step, 
starting from the initial one $n_\sigma(0)$. The undefined components of the self-energy at each final time can be accurately approximated as the self-energy one time step before i.e. 
$\Sigma_{HF,\sigma}^{\alpha\alpha}(t_N,t_N)\approx\Sigma_{HF,\sigma}^{\alpha\alpha}(t_{N-1},t_{N-1})$. The error introduced by this approximation becomes negligible for a sufficiently small $\Delta t$.
In the finite bandwidth situation this means $\Delta t\lesssim 1/D$ and in the wide band limit $\Delta t$ has to be taken smaller than the inverse of the greatest energy scale.
It can be checked that this procedure leads to the proper stationary values of $n_\sigma(t)$ in the unrestricted self-consistent HF approximation.\\

 In Fig. \ref{population_eq_U} we show the time evolution of the central level charge per spin at different levels of approximation. In Fig. \ref{population_eq_U} a) we compare the exact Monte Carlo (MC) 
results from Ref. 
\cite{Schmidt_PRB2008} with the ones obtained for the self-consistent and the non self-consistent (first order) HF approximation for the electron-hole symmetric situation ($\epsilon_0=-U/2$) and the 
$(n_\uparrow(0),n_\downarrow(0))=(0,0)$ initial configuration. As can be observed, the non self-consistent approximation tends to deviate from the exact results, leading to a stationary
charge overpassing the electron-hole symmetric stationary value. This result is in agreement with Ref. \cite{Schmidt_PRB2008}, where the authors analyzed the level population by means of a first order 
perturbation theory in the Coulomb interaction parameter $U/\Gamma$. Although a good agreement is found for small values of $U/\Gamma$, the charge progressively deviates from the exact results for 
increasing $U/\Gamma$. This pathological behavior is corrected within a fully self-consistent HF treatment, where the average charge per spin $n_\sigma(t)$ tends to the correct 
singlet state for all $U/\Gamma$ values. As shown below, inclusion of correlations provided by the second order diagrams further improve the agreement with 
the numerically exact results.\\

In Fig. \ref{population_eq_U} b) we show the level population evolution for an initially trapped spin, $(n_\uparrow(0),n_\downarrow(0))=(1,0)$. We have chosen a case with electron-hole symmetry 
($\epsilon_0=-U/2$) and with parameters such that $U/\pi\Gamma>1$, which leads to a magnetic solution in the stationary case within the HF approximation
\cite{Anderson_PR1961}. As can be observed, the numerical solution is in remarkable agreement with the exact expression of Eq. (\ref{charge_HF}). Let us comment that for initial conditions with unbroken 
spin symmetry, i.e. $(n_\uparrow(0),n_\downarrow(0))=(0,0)$, $(1,1)$, the system always tends to a non-magnetic solution for all values of $U/\Gamma$.\\

Finally, it is worth remarking that the prediction of a magnetic solution within the HF approximation at zero-temperature is well known to be unphysical as the ground state of the system should be always 
a singlet \cite{Wiegmann_PL1980,Kawakami_PLA1981,Andrei_RMP1983}. This behavior should be corrected when including electronic correlations in an appropriate way. In the next section we will analyze the 
effect of correlations beyond the HF approximation in the transient regime.\\

\subsection{Effects of correlation beyond the Hartree-Fock approximation}
\label{subsec::correlations_U}
Within a Green functions approach, correlation effects are included in the electron self-energy. In a stationary situation an appropriate second-order self-energy in the interaction 
parameter $U/\Gamma$ can include these effects in a rather satisfactory way. Indeed it can be shown that the exact self-energy in the limit $U/\Gamma\to\infty$ has a functional form close to the 
second order one and is in fact proportional to $U^2$ \cite{Martin_SSC1982}.
This fact has been used in different interpolative approaches based on the second-order self-energy giving a reasonable approximation for the 
Anderson model between the weak and strong coupling limits \cite{Martin_SSC1982,Martin_PRB1986,Yeyati_PRL1993,Kajueter_PRL1996}.\\

We will concentrate in the symmetric case, $\epsilon_0=-U/2$, where correlations effects are more important. It can be shown that the inclusion of the second-order self-energy yields a spectral density
in the equilibrium stationary case in rather good agreement with numerical renormalization group (NRG) calculations \cite{Anders_JCM2008}. Indeed in this approximation the charge peaks in the
spectral density are well described, fulfilling the Friedel
sum rule at zero energy, although somewhat overestimating the width of the Kondo resonance at very large $U/\Gamma$. It is important to notice that the second-order self-energy diagram has to be 
calculated with propagators including the HF correction to the energy level (i.e. the HF propagators) in order to ensure electron-hole symmetry. On the other hand, it can be shown that a fully 
self-consistent calculation of the diagrams (i.e. using fully dressed propagators) yields instead a poor description of the spectral density \cite{White_PRB1992}.\\

In a general time-dependent non-equilibrium situation the self-energy diagrams must be calculated in Keldysh space. The $+-$ ($-+$) components of the second order self-energy have the simple 
expressions
 \begin{eqnarray}
\hat{\Sigma}_{\sigma}^{(2)+-}(t,t')=-U^2 \hat{G}_{HF,\sigma}^{+-}(t,t')\hat{G}_{HF,\bar{\sigma}}^{+-}(t,t')\hat{G}_{HF,\bar{\sigma}}^{-+}(t',t),\nonumber\\
\hat{\Sigma}_{\sigma}^{(2)-+}(t,t')=-U^2 \hat{G}_{HF,\sigma}^{-+}(t,t')\hat{G}_{HF,\bar{\sigma}}^{-+}(t,t')\hat{G}_{HF,\bar{\sigma}}^{+-}(t',t),
\end{eqnarray}
where the HF propagators are calculated as indicated in Eq. (\ref{G_HF_U}). The other Keldysh components are then given by the usual Keldysh relations, making the same choice for equal times as in Eq. 
(\ref{half_criteria}). The propagators in Keldysh space can now be evaluated inverting Eq. (\ref{Dyson_discretized}) with $\hat{\Sigma}_{int,\sigma}=\hat{\Sigma}_{HF,\sigma}+\hat{\Sigma}^{(2)}_\sigma$.\\


\begin{figure}
   \begin{minipage}{0.49\textwidth}
     \centering
     \includegraphics[width=1\linewidth]{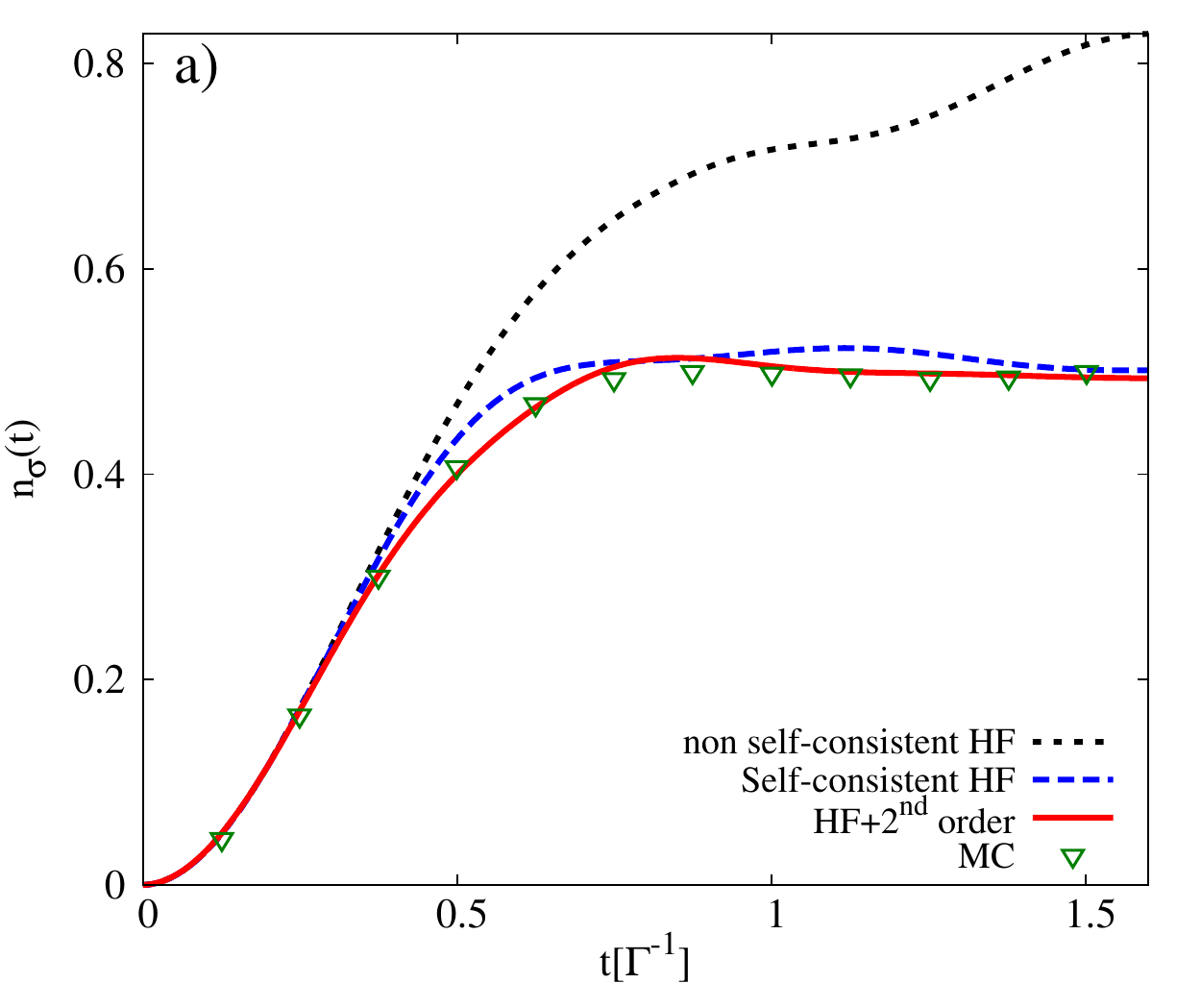}
   \end{minipage}\hfill
   \begin {minipage}{0.49\textwidth}
     \centering
     \includegraphics[width=1\linewidth]{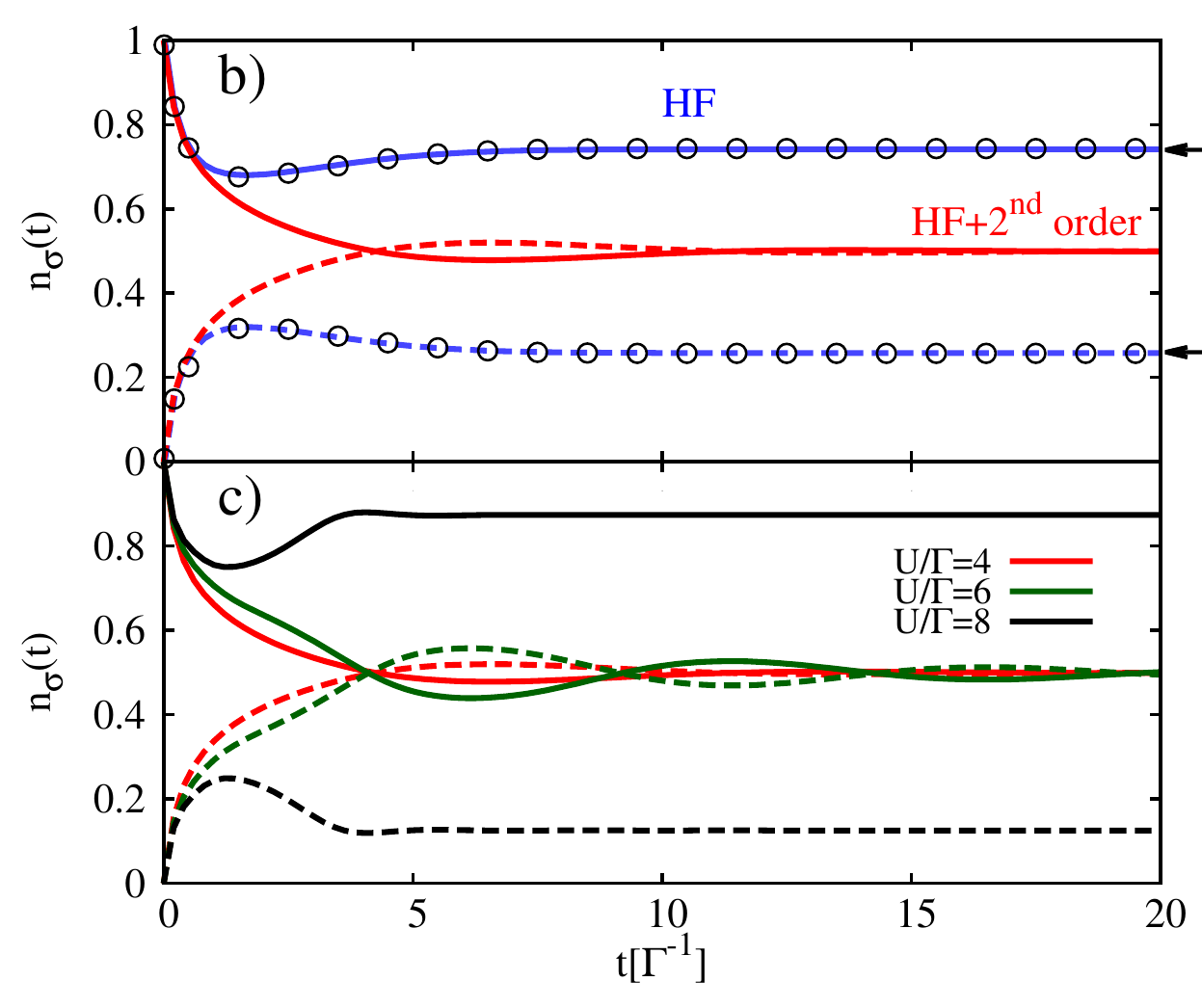}
   \end{minipage}
\caption{Time evolution of the level charge for the Anderson model. a): Average population per spin at different levels of approximation compared to MC simulations from 
Ref. \cite{Schmidt_PRB2008}, 
with $U/\Gamma=8$, $V=0$, $\epsilon_0=-U/2$, $T=0.2\Gamma$, $D=10\Gamma$ and $(n_{\uparrow}(0),n_{\downarrow}(0))=(0,0)$ initial configuration. b) and c): Results for the 
up (solid lines) and down (dashed lines) spin and the initial configuration $(1,0)$. In b) the HF approximation (blue lines) is compared to the analytic 
expression (black points), given by Eq. (\ref{charge_HF}) for $U/\Gamma=4$, $\epsilon_0=-U/2$ and in the infinite bandwidth limit. The arrows denote the stationary solution. The red lines correspond to 
the second-order self-energy case. In c) the level charge for the same parameters and three different Coulomb interactions $U/\Gamma=4$ (red), $6$ (green) and $8$ (black) is 
shown for the second order approximation. In continuous line we show the evolution of the spin up and in dashed the evolution of the down spin.}
    \label{population_eq_U}
\end{figure}

We will now analyze the effect of correlations on the electronic and transport properties of the system. In Fig. \ref{population_eq_U} a) we show the population evolution for the case discussed in the 
previous section and an initial configuration $(n_{\uparrow}(0),n_{\downarrow}(0))=(0,0)$. As can be observed, the inclusion of electron correlation effects improve the agreement with the exact MC 
calculations.\\

In  Fig. \ref{population_eq_U} b) we show the evolution of $n_\sigma(t)$ with an initial configuration $(n_\uparrow(0),n_\downarrow(0))=(1,0)$ in which a magnetic solution was predicted by the HF 
approximation. As it can be observed, 
when including correlations (electron-hole pair creation) the system evolves to a non-magnetic solution corresponding to a singlet state in the stationary limit.
In Fig. \ref{population_eq_U} c) we analyze the evolution of $n_\sigma(t)$ for the same initial magnetic configuration for increasing values of the electron-electron interaction 
parameter. It is found that for $U/\Gamma\gtrsim8$ the initial localized spin is no longer screened by the electrodes, tending to a magnetic solution. This indicates a shortcoming of the approximate 
self-energy for sufficiently large interaction strength. The singlet stationary state is, however, always reached when starting from a configuration without spin-symmetry breaking.\\

\begin{figure}
   \begin{minipage}{0.49\textwidth}
     \centering
     \includegraphics[width=1.0\linewidth]{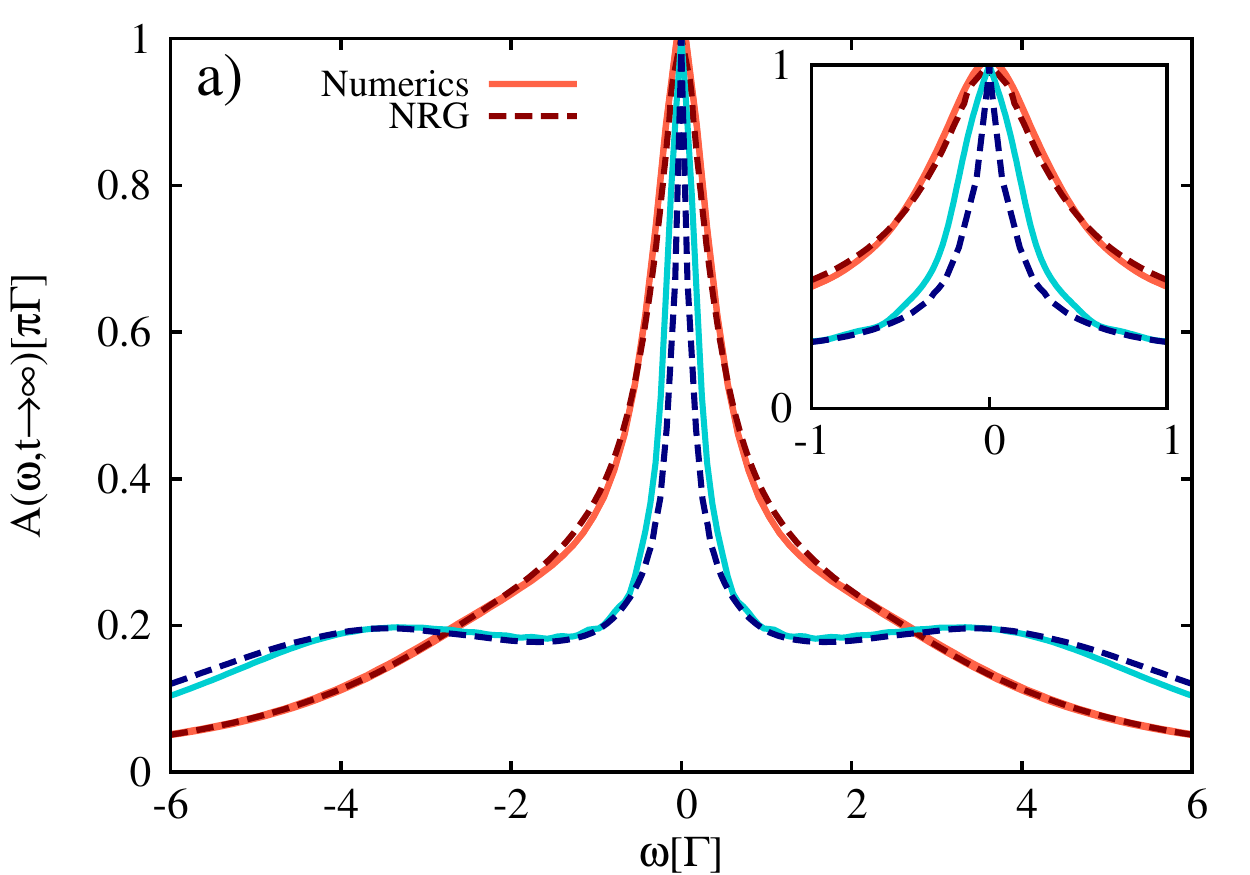}
   \end{minipage}
   \begin {minipage}{0.5\textwidth}
     \centering
     \includegraphics[width=1.0\linewidth]{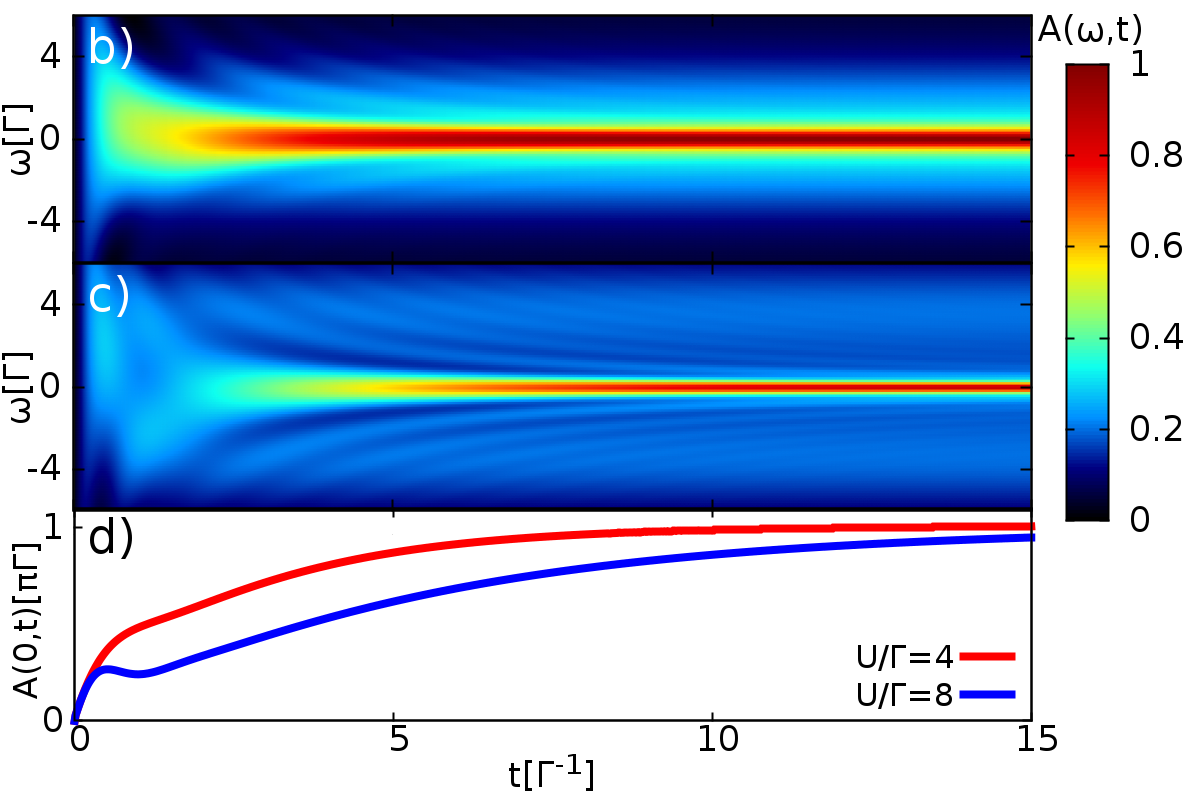}
   \end{minipage}
\caption{a): Long time DOS for different electron-electron interactions, $U/\Gamma=4$ (red curves) and $8$ (blue curves) and $V=0$. We compare
the results of the perturbation expansion up to second order (full lines) with those from 
the exact NRG calculation of Ref. \cite{Anders_JCM2008} (dashed lines). The inset show the convergence of the Kondo resonance. b) and c): Time evolution of the density of 
states, showing the formation of the Kondo peak for $U/\Gamma=4$ and $U/\Gamma=8$, respectively. In d) the height of the Kondo peak is represented as a function of time
 for these two cases.}
    \label{spectral_U}
\end{figure}

In Fig. \ref{spectral_U} a) we analyze now the long time evolution of the DOS, $A(\omega,t\to\infty)$.  These results demonstrate 
that the second order self-energy provides a good approximation to the problem \cite{Han_PRL2007}, leading to a remarkable agreement with results from 
NRG calculations for moderate $U/\Gamma$ values
\cite{Anders_JCM2008}. The inset in this panel shows a blow up of the Kondo resonance, where it can be observed that the second order self-energy tends to overestimate its width for large $U/\Gamma$ values.\\


It should be remarked that the convergence time increases with $U/\Gamma$. In this respect it is interesting to analyze the convergence in time of the Kondo resonance, an issue 
that has been addressed in previous works \cite{Norlander_PRL1999,Nghiem_PRL2017}. One would expect this convergence time to be of the order of $T_{K}^{-1}$, $T_K$ being the Kondo temperature. In Figs. \ref{spectral_U} b) and c) we show the time evolution of the spectral density for two values of the interaction strength, $U/\Gamma=4$ and $8$. The formation in time of the Kondo resonance 
is illustrated, showing a longer time for the larger interaction. Considering the expression for the Kondo temperature in the electron-hole symmetric Anderson model, i.e.
$T_K = \sqrt{U\Gamma/2} \exp[-\pi U/8\Gamma]$, for these cases we have the ratio $T_K(U/\Gamma=4)/T_K(U/\Gamma=8) \simeq 3.4$. Thus, one would expect
a Kondo resonance formation time for the $U/\Gamma=8$ case roughly $3.4$ times larger than for $U/\Gamma=4$. The ratio of formation times that can be estimated from Figs. 
\ref{spectral_U} b) and c) is somewhat smaller due to the slight overestimation of the width of the Kondo peak by the second order diagrammatic 
approximation for the larger $U/\Gamma$ value. On the other hand, Fig. 
\ref{spectral_U} d) shows the evolution of the height of the central peak, $A(\omega=0,t)$, to its stationary value fixed by the Friedel sum rule 
$A(\omega=0,t\to\infty)=1/\pi\Gamma$. 
A kink in the evolution is observed at times $\sim1/U$, mainly visible for large $U/\Gamma$ values, due to the appearance of the charge bands.\\

\begin{figure}
   \begin{minipage}{0.49\textwidth}
     \centering
     \includegraphics[width=1\linewidth]{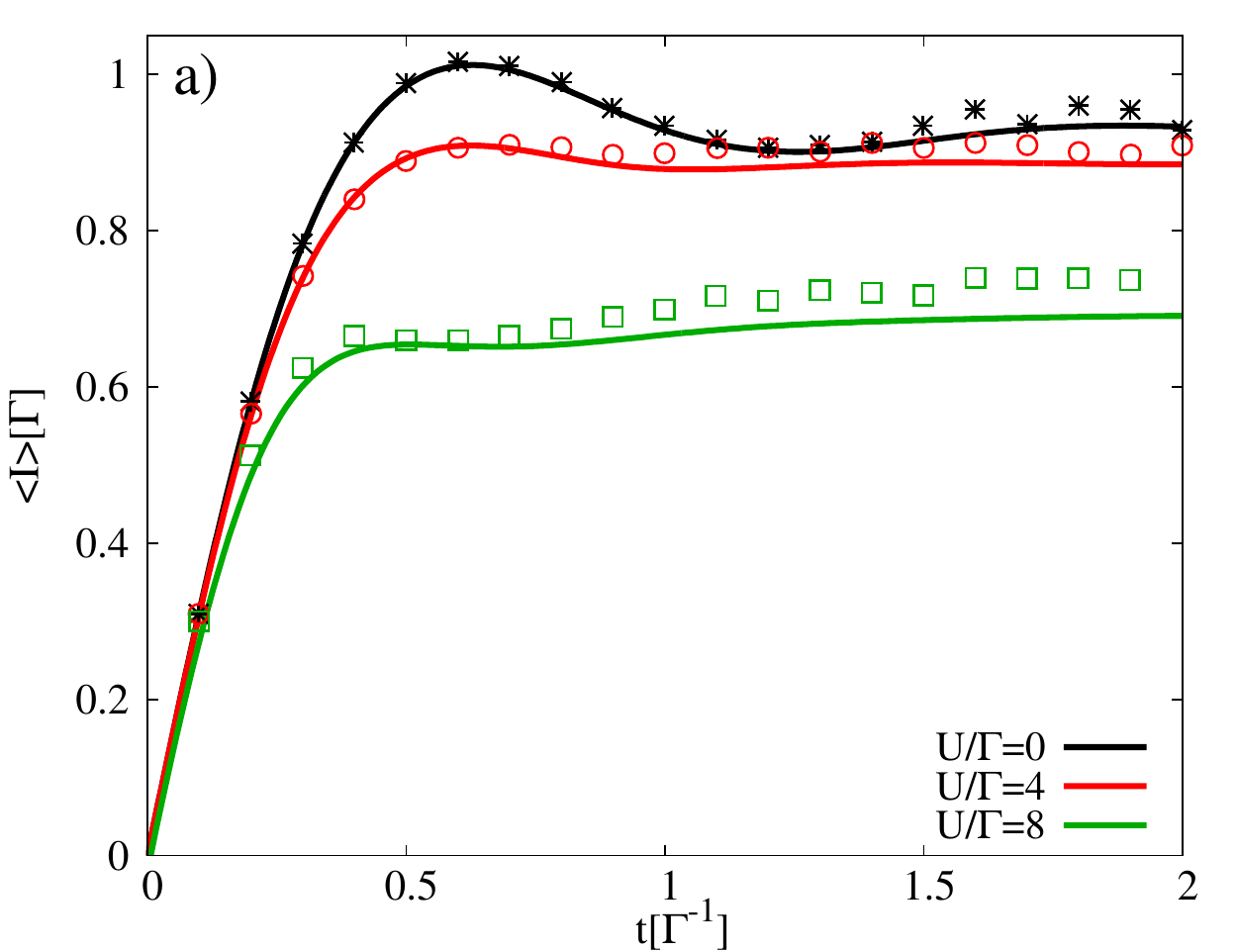}
   \end{minipage}\hfill
   \begin {minipage}{0.49\textwidth}
     \centering
     \includegraphics[width=1\linewidth]{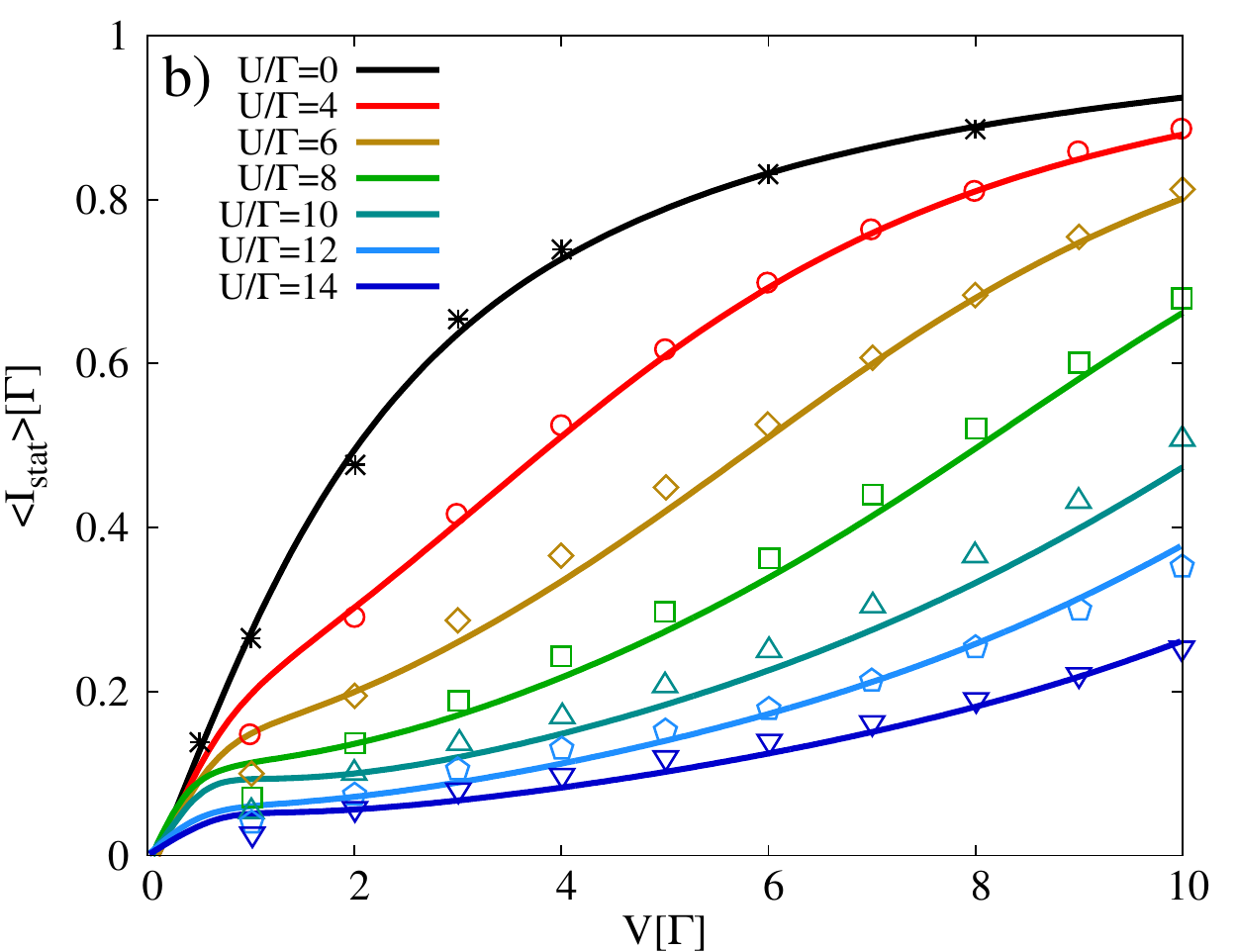}
   \end{minipage}
\caption{a): Short time symmetrized current ($\langle I\rangle=(I_L-I_R)/2$), comparing the results for the perturbation expansion up to second order
(solid lines) with the ones obtained using MC in Ref. \cite{Schmidt_PRB2008} (symbols)
for $U/\Gamma=0$, $4$ and $8$, with $V=10\Gamma$, $\epsilon_0=-U/2$, $T=0.2\Gamma$ and $D=10\Gamma$. The asymptotic current as a function of the voltage is shown in b) for 
increasing values of the electron-electron interaction, compared with the exact MC results from Ref. \cite{Werner_PRB2009} (symbols).}
    \label{current_U}
\end{figure}

Let us discuss now the voltage biased situation. In Fig. \ref{current_U} a) we show the evolution of the current for the second order perturbation expansion
together with results from the MC simulations
 finding also a good quantitative agreement. For very large interaction strengths the agreement becomes somewhat poorer although still capturing the general trend.\\


Finally in Fig. \ref{current_U} b) we show the asymptotic $I(V)$ characteristic 
for increasing $U/\Gamma$ values compared to the MC results of Ref. \cite{Werner_PRB2009}. As can be observed, there is an overall good agreement specially for $V>\Gamma$. 
However, for $V < \Gamma$ the second order self-energy tends to
slightly overestimate the current due to the already mentioned shortcoming in the description
of the Kondo resonance. In fact, this approximation is unable to describe the
splitting of this resonance for $V <T_K$. This shortcoming would be removed
in this electron-hole symmetric case by including the fourth order diagrams, as shown in Ref. \cite{Fujii_PRB2003} in the stationary limit.\\

\section{Electron-phonon interaction: Spinless Anderson-Holstein model}
\label{sec::eph}
In order to analyze the transient regime in the presence of electron-phonon interactions we will first consider the spinless Anderson-Holstein model \cite{Holstein_model_59}. In this model an electron in 
the central level is coupled to a single vibrational mode. The Hamiltonian of the system is given by
\begin{equation}
 \hat{H}=\hat{H_0}+\hat{H}_{ph}+\hat{H}_{e-ph}\;,
\label{H_e-ph}
\end{equation}
where $\hat{H}_{0}$ is the non-interacting part in the Hamiltonian of Sec. \ref{sec::formalism}, $\hat{H}_{ph}=\omega_0 b^\dagger b$, $\omega_0$ being the frequency of the local phonon mode
 and $b$ ($b^\dagger$) the phonon annihilation (creation) operator. The electron-phonon interaction at the central region is described by the term $\hat{H}_{e-ph}=\lambda(b^\dagger+b)d^\dagger d$, 
where $\lambda$ measures the electron-phonon coupling strength.\\

\subsection{Hartree solution}
\label{subsec::HF_lambda}
\begin{figure}
\includegraphics[width=.9\linewidth]{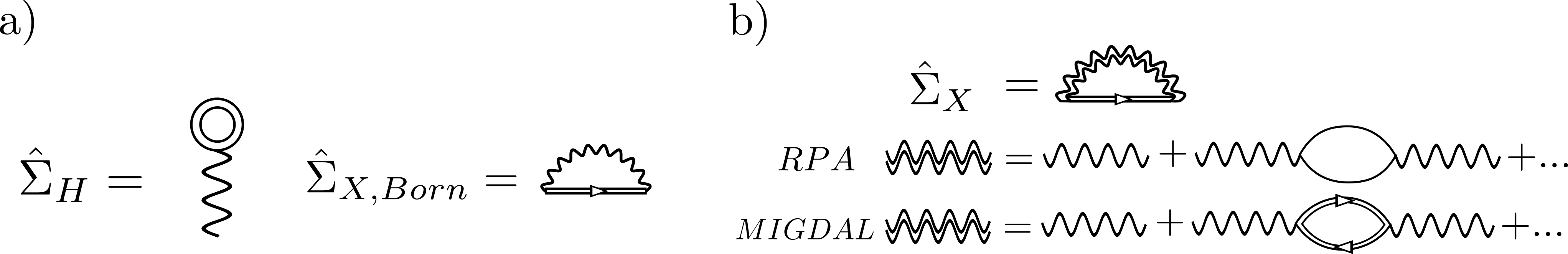}
\caption{Second order self-energy diagrams for the spinless Anderson-Holstein interaction. a): diagrams for the Born approximation, using the bare phonon propagator (wavy line). 
In b) similar approximations are shown with two different schemes for dressing the phonon propagator: RPA \cite{Utsumi_PRB2013}, where the electronic propagators are considered to 
be undressed, and the self-consistent MIGDAL \cite{Murakami_PRB2015}, where the electronic propagators are fully dressed.}
\label{Diagrams_L}
\end{figure}

As in the previous section, we begin our analysis with the self-consistent mean-field approximation in which the self-energy is approximated by the ``tadpole'' diagram of Fig. \ref{Diagrams_L}
(Hartree approximation). Within this approximation, the self-energy in Keldysh space can be evaluated as
\begin{eqnarray}
 \Sigma_{H}^{\alpha\alpha}(t,t')=\alpha\delta(t-t')\lambda^2\int d\tau\left[d^{++}(t,\tau)-d^{+-}(t,\tau)\right]n(\tau)\,,\qquad
 \Sigma_{H}^{+-}(t,t')=\Sigma_{H}^{-+}(t,t')=0\,,
 \label{S_phon_HF}
\end{eqnarray}
where $n(t)$ is the self-consistent central level charge and $\hat{d}$ is the free phonon propagator in Keldysh space given by
\begin{equation}
\hat{d}(t,t') = -i \left(\begin{array}{cc}2n_p\cos\omega_0(t-t')+e^{-i\omega_0|t-t'|} &\, n_pe^{-i\omega_0(t-t')}+(n_p+1)e^{i\omega_0(t-t')}\\ 
n_pe^{i\omega_0(t-t')}+(n_p+1)e^{-i\omega_0(t-t')} &\, 2n_p\cos\omega_0(t-t')+e^{i\omega_0|t-t'|} \end{array}\right) \,,
\label{phonon_prop}
\end{equation}
where $n_p=(e^{\omega_0/T}-1)^{-1}$ is the free phonon population, described in a thermal equilibrium situation by the Bose-Einstein distribution. Most of the calculations are performed at zero or very small temperature, considering $n_p=0$.
Using the Keldysh relations, Eqs. (\ref{S_phon_HF}) can then be written as
\begin{equation}
 \Sigma_{H}^{\alpha\beta}(t,t')=\alpha\lambda^2\delta(t-t')\delta_{\alpha\beta}\int_{0}^{t} d\tau d^R(t,\tau)n(\tau)\,,
\label{Holstein_HF}
\end{equation}
where $d^R(t,t')$ is the retarded free phonon propagator
\begin{equation}
 d^R(t,t')=-2\theta(t)\theta(t-t')\sin\left[\omega_0(t-t')\right]\,.
\end{equation}
It is worth noticing that, at variance with the case of the electron-electron interaction discussed in the previous section, the electron-phonon interaction introduces retardation effects even in 
the Hartree approximation. These effects will be important in the transient regime except in the limit of a sufficiently fast phonon ($\omega_0\gg\epsilon_0,\Gamma$) \cite{Riwar_Schmidt} with a central 
charge evolving adiabatically. In this limit Eq. (\ref{Holstein_HF}) tends to 
\begin{equation}
 \Sigma_{H}^{\alpha\beta}(t,t') \approx-\alpha\delta(t-t')\delta_{\alpha\beta}\frac{2\lambda^2}{\omega_0}n(t)\,.
 \label{Holstein_HF_inst}
 \end{equation}

\begin{figure}
   \begin{minipage}{0.49\textwidth}
     \centering
     \includegraphics[width=1\linewidth]{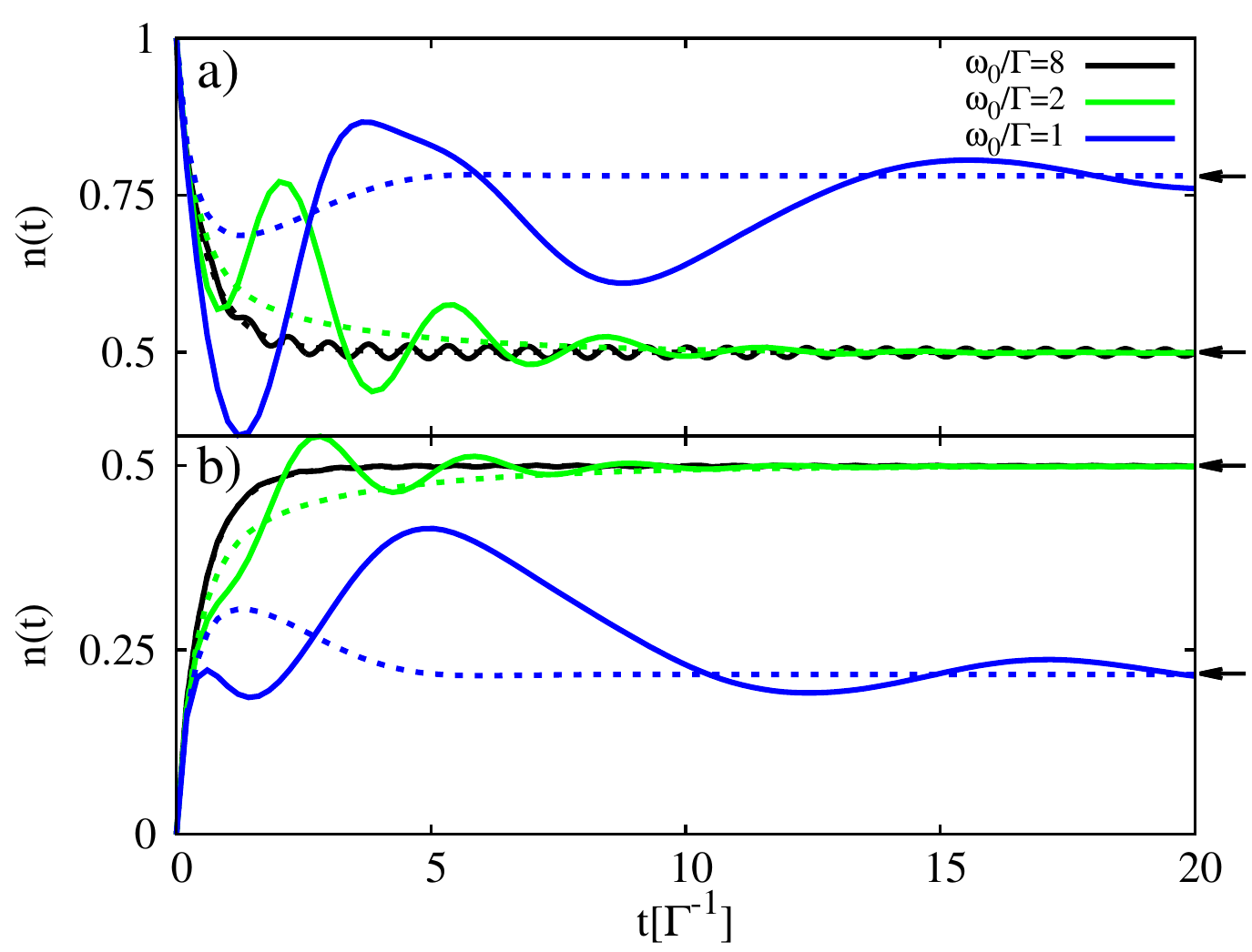}
   \end{minipage}\hfill
   \begin {minipage}{0.49\textwidth}
     \centering
     \includegraphics[width=1\linewidth]{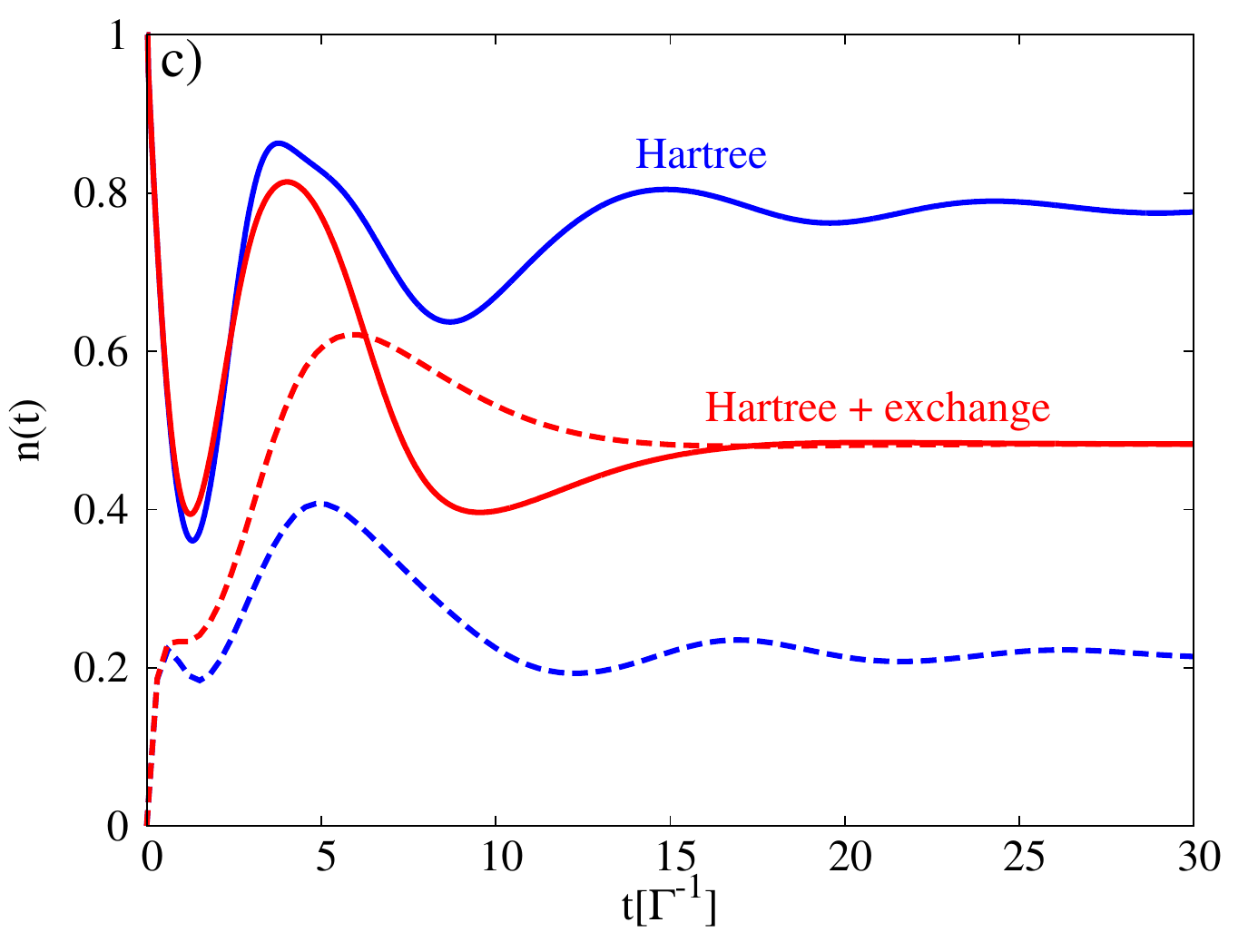}
   \end{minipage}
\caption{Time evolution of the central level charge for an initially full in a), and empty level in b). The dotted lines represent the evolution using an instantaneous Hartree term Eq. 
(\ref{Holstein_HF_inst}), while the solid ones correspond to the full Hartree self-energy Eq. (\ref{Holstein_HF}). The dependence on phonon frequency is also illustrated for
 the values: $\omega_0=8\Gamma$ (black), $2\Gamma$ (green) and $\Gamma$ (blue). c): Charge evolution for an initially empty (dashed line) and initially full level (solid line) for the
 $\omega_0=\Gamma$ case. The blue and red lines correspond to the Hartree and self-consistent Born approximation, respectively. The remaining parameters are $\lambda=1.5\Gamma$, $V=0$ and the central 
level is set to $\epsilon_0=\lambda^2/\omega_0$, thus preserving electron-hole symmetry.}
    \label{population_eq_L}
\end{figure}

 We can now follow a similar procedure to the one used in the previous section to calculate $\hat{G}_{H}$ and the central level self-consistent charge. Figs. \ref{population_eq_L} 
a) and b) show the evolution of the level charge in the transient regime. As in the case of electron-electron interactions, the charge evolves to the
 stationary value, indicated by the arrows in the figure. Figs. \ref{population_eq_L} a) and b) also illustrate how the solution progressively deviates from the adiabatic
 approximation given by Eq. (\ref{Holstein_HF_inst}) when reducing the value of $\omega_0$. 
The full self-consistent solution as given by the self-energy in Eq. (\ref{Holstein_HF}),  describes 
the time-dependent modification of the central level charge at time $t$ induced by
its past history at time $\tau<t$. Retardation effects of the phonon dynamics results in 
a coherent superposition of oscillations with period $2\pi/\omega_0$ but with different amplitudes ($\propto n(\tau)$). 
In the intermediate regime where the electron and the phonon dynamics are equally fast ($\omega_0\approx\Gamma$), the coherence between those oscillations is lost 
at long times ($t \gg 1/\Gamma,2\pi/\omega_0$), thus resulting in an effective damping of the central level charge, see Figs. \ref{population_eq_L} a) and b).
However, in the adiabatic regime ($\omega_0\gg\Gamma$) the dynamics of the electrons and phonons decouple, and small charge oscillations
persist on time, mostly in the $n(0)=1$ case (black curve in Fig. \ref{population_eq_L} a). A natural lifetime describing the decay of those oscillations could be included by 
dressing the phonon line in the Hartree term depicted in Fig. \ref{Diagrams_L} b).\\

 Finally, one can observe in Figs. \ref{population_eq_L} a) and b) that for the smallest values of $\omega_0$ two
 different asymptotic charge values are reached depending on the initial level population. This is the charge bistable behavior predicted by the self-consistent Hartree approximation
 in the strong-coupling limit \cite{Damico_NJP2008,Riwar_Schmidt}. For the case of electron-hole symmetry considered in Fig. \ref{population_eq_L} and at zero temperature and bias voltage,
the condition for the appearance of bistability is $2\lambda^2/\pi\Gamma\omega_0 >1$. The possibility of a bistable regime for a molecular quantum dot with strong electron-phonon interaction was
 suggested some time ago \cite{Gogolin_2002,Alexandrov_PRB2003,Galperin_nano2005}. 
The interest in investigating such a phenomenon has experienced a recent revival. For instance, it has been shown 
that the displacement fluctuation spectrum of a nanomechanical oscillator strongly coupled to electronic transport, either 
in the regime of semiclassical phonons \cite{Micchi_PRL2015,Micchi_PRB2016}, 
or for a quantum nanomechanical oscillator entering the
Franck-Condon regime \cite{Avriller_PRB2018} bears clear signatures of a transition to a bistable regime.
Moreover, by making a mapping to the Kondo problem,
the bistability was shown to be destroyed in equilibrium conditions by quantum fluctuations if the temperature 
is lower than a phonon mediated Kondo temperature \cite{Klatt_PRB2015}.
Notice, however, that this phonon displacement bistability does not correspond necessarily to a bistable behavior for the charge or the current, as predicted
by the mean field approximation. As even this simple spinless Anderson-Holstein model is not exactly solvable, this issue
is still under debate \cite{Albrecht_PRB2012,Wilner_PRB2013,Albrecht_PRB2013}. It seems to us plausible that, at 
least for equilibrium conditions and $T=0$ correlation effects destroy the charge and current bistability predicted by the mean field solution. We address this issue in the following section.


\subsection{Effects of correlation beyond Hartree approximation}
\label{subsec::S2_lambda}
We will go beyond the mean-field solution by analyzing three different approximations for the self-energy. We first consider the self-consistent Born approximation given by the diagrams in Fig.
\ref{Diagrams_L} a) This is a conserving approximation in which the diagrams are calculated from the fully dressed electron propagators. The phonon propagator is however not renormalized. Within this
approximation both diagrams appearing in Fig. \ref{Diagrams_L} a) have the expression
\begin{eqnarray}
 \hat{\Sigma}^{\alpha\beta}_{H}(t,t')=-2\lambda^2\alpha\delta_{\alpha\beta}\delta(t-t')\int_{0}^td\tau\sin\left[\omega_0(t-\tau)\right]n(\tau)\,,\qquad
 \hat{\Sigma}^{\alpha\beta}_{X}(t,t')=i\alpha\beta\lambda^2\hat{G}^{\alpha\beta}(t,t')\hat{d}^{\alpha\beta}(t,t'),
 \label{sigma_holstein}
\end{eqnarray}
where $\hat{G}^{\alpha\beta}$ denotes the Keldysh components of the
fully dressed electron propagators and $n(t)$ is the final self-consistent charge.\\


This fully self-consistent approximation can be straightforwardly implemented within the numerical procedure of Sec. \ref{sec::formalism}. For each time in the discretized mesh, the self-energies of Eqs. 
(\ref{sigma_holstein}) are calculated from the final Green functions and then stored. As in the case of the Hartree solution previously discussed, when inverting Eq. (\ref{Dyson_discretized}) for each 
time the self-energies at the final time in each iteration are not well defined but its value can be extrapolated from the ones calculated
at the previous mesh point in the time grid.
For sufficiently small $\Delta t$ the 
error introduced by this approximation becomes negligible. We have 
checked the accuracy of this procedure by verifying that the solution tends to the proper stationary one.\\

In Fig. \ref{population_eq_L} c) we show the evolution of the central level charge for a choice of parameters in which the Hartree approximation predicts a bistable behavior. 
As can be observed, the inclusion of
correlations eliminates the charge bistability appearing in the Hartree approximation, tending to the same asymptotic value for the initially empty and full cases. We have checked that this behavior is 
maintained up to quite
large values of $\lambda^2/\omega_0\Gamma$, although eventually the self-consistent Born approximation breaks down in the strong polaronic regime. This indicates that another kind of approximation has to
be used to explore this parameter regime, like for instance in the lines of the ones discussed in Refs. \cite{Martin_PRB2008,Maier_PRB2011,Albrecht_PRB2013,Dong_PRB2013,Souto_PRB2014}. 
These results suggest that the bistable behavior of the central level charge predicted in Refs. \cite{Damico_NJP2008,Riwar_Schmidt} is a spurious feature of the mean field approximation which disappears
when correlation effects are included. This is in agreement with the predictions of exact numerical calculations of Ref. \cite{Klatt_PRB2015}, at least for the equilibrium case and at sufficiently low 
temperatures. It does not imply that an apparent bistability might not be observed for a continuous bath model or adopting more general initial conditions for the phonon mode density matrix 
\cite{Wilner_PRB2014,Wilner_PRB2015}.\\

So far, the renormalization of the phonon propagator has been neglected. The simplest way to include this effect is by means of an RPA-like approximation \cite{Utsumi_PRB2013}. 
The phonon propagator will satisfy a Dyson equation in Keldysh space similar to the electronic one given in Eq. (\ref{Dyson})
\begin{equation}
 \hat{D}=\left(\hat{d}^{-1}-\hat{\Pi}\right)^{-1}\,,
\label{Dyson_phon}
\end{equation}
where $\hat{D}(t,t')=-i\left\langle \hat{T}_c[\hat{\varphi}(t)\hat{\varphi}^{\dagger}(t')]\right\rangle$, with $\hat{\varphi}=b+b^\dagger$. $\hat{d}^{-1}$ is the inverse free-phonon propagator and 
$\hat{\Pi}$ is 
the phonon self-energy given by
\begin{equation}
 \hat{\Pi}^{\alpha\beta}(t,t')=-i\alpha\beta\lambda^2G^{\alpha\beta}(t,t')G^{\beta\alpha}(t',t)\,.
\end{equation}
As in the electronic case, Eq. (\ref{Dyson_phon}) can be discretized in a time mesh along the Keldysh contour. In order to solve numerically the corresponding matrix equation, an 
expression for the inverse free phonon propagator discretized on the contour must be obtained. This is a task which, to best of our knowledge, has not been achieved in
the literature, the mathematical difficulty lying in the fact that the inverse phonon propagator becomes singular in the free limit. This singularity must be then
somehow regularized. To obtain an expression of $\hat{d}^{-1}$ we have developed a regularization procedure which is discussed in Appendix \ref{App::d_0}, finding
\begin{equation}
{\bold{d}}^{-1} = \frac{1}{2\delta}\left(\begin{array}{ccccc|ccccc} 
h^{+}_0 & -1 & & & & & & & &h_{0N} \\
-1 & h & -1 & & & & & & & \\
 & \ddots &\ddots &\ddots & & & & & & \\
 & & -1& h &-1 & & & & &\\
 & & & -1 & h^{+}_N & c & & & &\\
\hline  
 & & & & c & h^{-}_N & 1 & & &\\
 & & & &  & 1 & -h & 1 & &\\
 & & & & & & \ddots & \ddots & \ddots &\\
 & & & & &  & & 1 &  -h & 1\\
h_{0N} & & & & & & & & 1 & h^{-}_0\\
 \end{array} \right)_{2N\times2N} \, ,
\label{kamenev_phon}
\end{equation}
where $\delta=\Delta t\,\omega_0$ and $h=2(1-\delta^2/2)$. The information about the initial phonon state is encoded in the components 
$h^{\pm}_0=\pm h/2 + i\delta(1+\rho_{0}^2)/(1-\rho_{0}^2)$ and $h_{0N}=-2i\delta\rho_0/(1-\rho_{0}^2)$, where $\rho_0=n_p(0)/[n_p(0)+1]$ and $n_p(0)$ is the initial phonon population. We will consider that 
phonons are initially in thermal equilibrium and thus $\rho_0=e^{-\omega_0/T}$. The regularization procedure requires introducing an infinitesimal quantity
$\eta$ which enters in the matrix elements connecting both Keldysh branches:
$c=-2i\delta/\eta$ and $h^{\pm}_N=\pm h/2 - c$. The parameter $\eta$ can be interpreted as a small phonon relaxation rate
which has to be taken such as $1/\eta\gg t,1/\omega_0$ for a good convergence to the
expected free propagator when inverting Eq. (\ref{kamenev_phon}).\\

It should be noticed that this problem with the inversion of the free phonon propagator has been avoided in the literature by neglecting fast oscillating terms of the type 
$\left\langle \hat{T}_c[\hat{b}(t)\hat{b}(t')]\right\rangle$ and $\left\langle \hat{T}_c[\hat{b}^{\dagger}(t)\hat{b}^{\dagger}(t')]\right\rangle$ in the diagrammatic
expansion of $\hat{D}$ . This corresponds to the
so-called rotating wave approximation, which describes the regime where the phonon timescale is much faster than the electron dynamics ($\omega_0\gg\Gamma,\lambda$) 
\cite{Meyer_PRL2002,Agarwalla_PRB2016}. For the calculation of the phonon self-energy, $\hat{\Pi}$, we will analyze two different approximations. In the first one (that will be denoted as RPA) the 
propagators in the electron ``bubble'' are the non-interacting ones, whereas the fully dressed propagators will be used in the second one (denoted as MIGDAL), see Fig. 
\ref{Diagrams_L} b).\\

\begin{figure}
\includegraphics[width=0.69\linewidth]{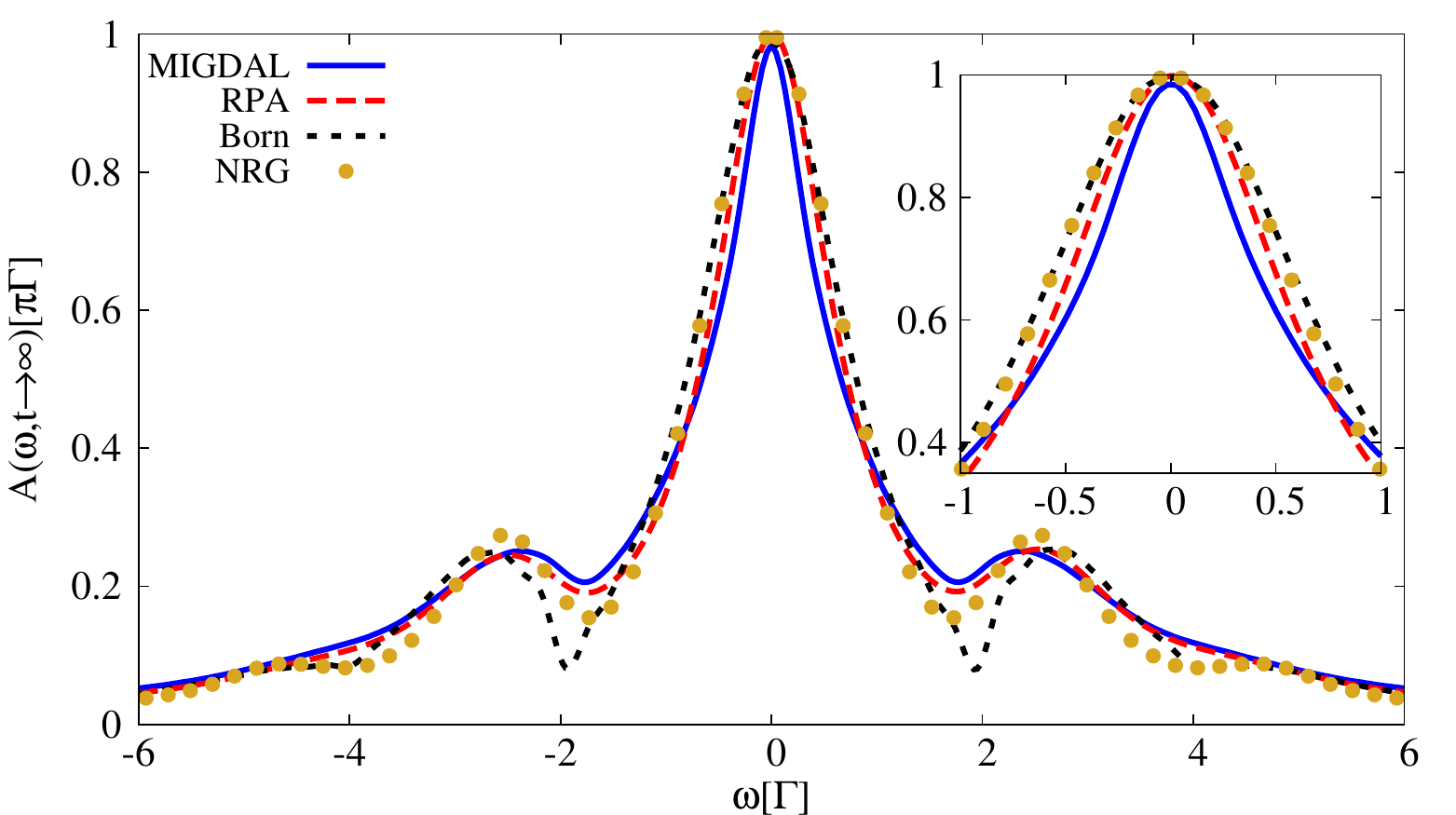}
\caption{Long time spectral density for the self-consistent MIGDAL (solid line), RPA (dashed) and Born (dotted) approximations, compared to NRG calculations (yellow dots). The inset shows the 
convergence of the central peak to the expected stationary value for the Born and RPA approximations. Parameters: $\lambda=1.5$, $\omega_0=2$, $V=0$, $\epsilon_0=\lambda^2/\omega_0$, $\Gamma=1$ and 
$D=30$.}
\label{spectral_RPA}
\end{figure}

In Fig. \ref{spectral_RPA} we show the long-time DOS at the central level for the three approximations considered in this section using the same parameters as in Fig. \ref{population_eq_L} with 
$\omega_0=2\Gamma$; a case with a rather strong electron-phonon coupling although still far from the polaronic limit ($\lambda^2/(\omega_0 \Gamma) \gg 1$). Notice the dip in the DOS at 
$\omega\approx\omega_0$ in the self-consistent Born approximation, which is a feature due to the logarithmic divergence of the second order self-energy $\hat{\Sigma}_X(\omega)$ at $\omega=\omega_0$ 
\cite{Martin_PRB2008,Laakso_NJP2014}. As it can be observed, both RPA and MIGDAL approximations, which include phonon renormalization eliminate 
this pathological divergence. A slight shift of the resonance around $\omega_0$ due to the renormalization of the phonon mode in both approximations can be observed. Notice 
also that all these approximations lead to an additional feature at $\sim2\omega_0$, associated to the appearance of a second phonon sideband. As an additional remark, in all cases the zero energy 
spectral density tends to reach the expected value predicted by the Friedel sum rule \cite{Jovchev_PRB2013}.\\

A further check of these approximations can be made by comparing their long-time DOS with the one predicted by a NRG calculation. To this end we have performed a NRG calculation of the stationary
DOS for the parameters of Fig. \ref{spectral_RPA}. As can be observed the agreement with the results of both RPA and MIGDAL is quite good for this parameter range. It should be commented that neither of 
these approximations are expected to be valid in the strong polaronic limit. Thus, features like the exponential decrease of the central resonance together with the appearance of a multiphonon structure 
in the DOS \cite{Martin_PRB2008,Maier_PRB2011,Albrecht_PRB2013,Dong_PRB2013,Souto_PRB2014,Martin_PRB2008} would require an approximation for the self-energy valid in the polaronic regime, as commented 
above.\\

\begin{figure}
   \begin{minipage}{0.49\textwidth}
     \centering
     \includegraphics[width=1\linewidth]{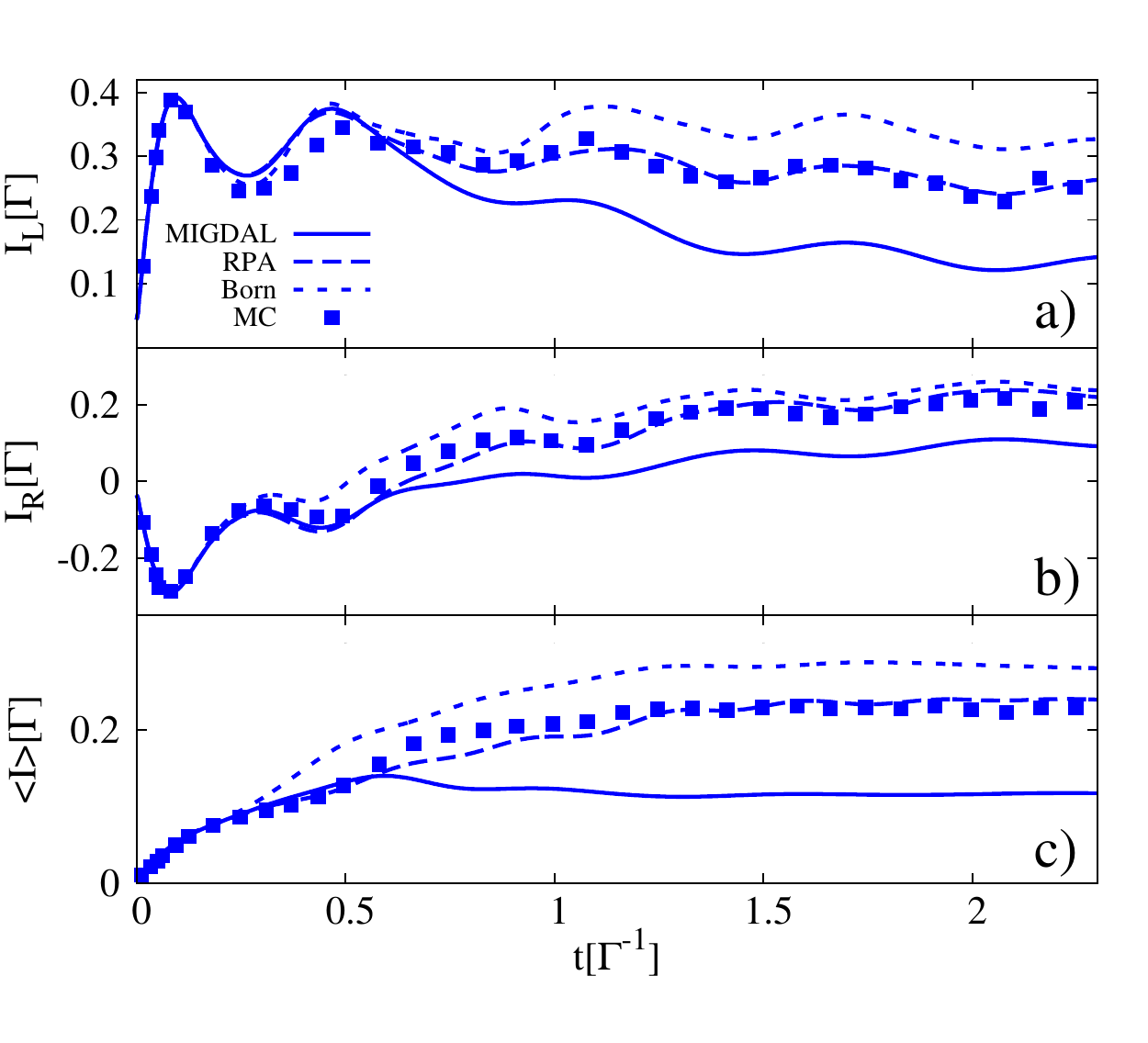}
   \end{minipage}\hfill
   \begin {minipage}{0.49\textwidth}
     \centering
     \includegraphics[width=1\linewidth]{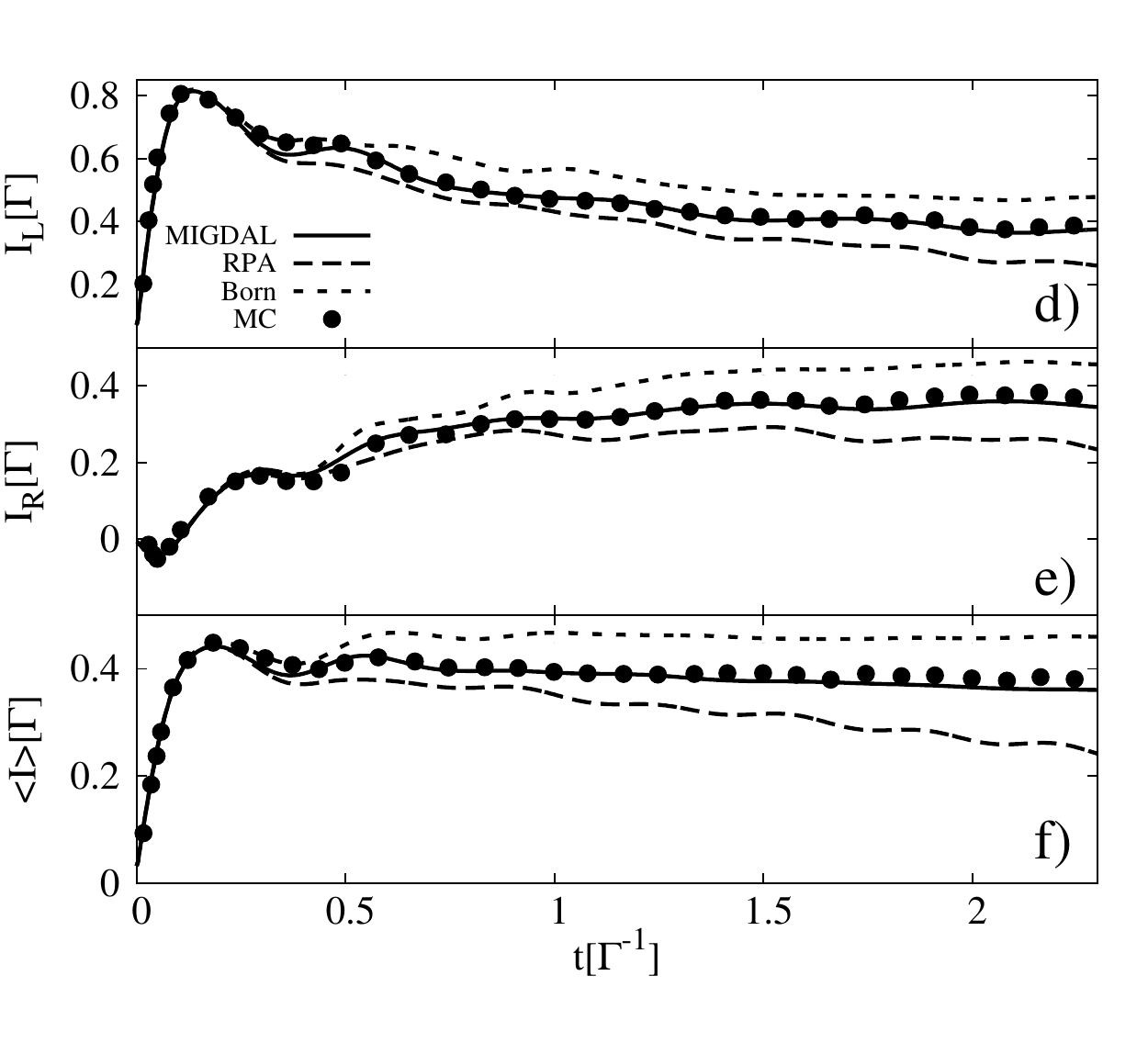}
   \end{minipage}
\caption{Comparison of the left a) and d), right b) and e) and symmetrized c) and f) current results with the MC simulations of ref. \cite{Muhlbacher_PRL2008}. Results for $V=4$, 
a)-c) and $32$ d)-f) cases are shown, and for the three approximations described in the text: self-consistent Born (dotted line), RPA (dashed line) and MIGDAL (solid line) approximations. 
The remaining parameters are: $\lambda=8$, $\omega_0=10$, $D=20$, $T=0.2$ and $\Gamma=1$.}
    \label{comparison_MC}
\end{figure}

Finally, in Fig. \ref{comparison_MC} we show results from the
three approximations for the transient left, right and average currents compared
to results obtained using MC simulations in Ref. \cite{Muhlbacher_PRL2008}. Both cases correspond to a rather strong interaction ($\lambda=8$, $\omega_0=10$ and
$\Gamma=1$) but two different bias voltages. Strikingly, as can be observed, 
RPA captures remarkably well the quantitative behavior of the numerically exact results
in the small voltage case, see Figs. \ref{comparison_MC} a)-c), whereas for very large
bias it is the MIGDAL approximation that gives a better quantitative agreement
with the MC numerical results, see Figs. \ref{comparison_MC} d)-f). This would indicate that the inclusion
of phonon renormalization and non-equilibrium effects (like heating of the local vibrational mode under increasing bias voltage) are essential for a good 
description of this regime. Furthermore, the higher the bias voltage the better
this effects are included in the fully self-consistent approach given by MIGDAL.

\section{Electron-electron and electron-phonon interactions}
\label{sec::AH}
In this section we study the transient regime in the presence of both electron-electron and electron-phonon interactions. We consider the
spin-degenerate Anderson-Holstein model defined as
\begin{equation}
 \hat{H}=\sum_{\sigma=\uparrow,\downarrow}\hat{H}_{0,\sigma}+\hat{H}_{e-e}+\hat{H}_{ph}+\hat{H}_{e-ph}
\end{equation}
where $\hat{H}_{0,\sigma}$ is the non-interacting part of the Hamiltonian 
given in Sec. \ref{sec::formalism}, $\hat{H}_{e-e}=U\hat{n}_\uparrow \hat{n}_\downarrow$,  $\hat{H}_{ph}=\omega_0b^\dagger b$ and 
$\hat{H}_{e-ph}=\lambda(b+b^\dagger)\sum_{\sigma} \hat{n}_{\sigma}$.
In this case we combine the approximations used in Sec. \ref{sec::eph} for the electron-electron
self-energies with the ones in the previous section for the electron-phonon case,
i.e. $\Sigma_{int} = \Sigma_{e-e} + \Sigma_{e-ph}$
(see Figs. \ref{contour_Diagrams_U} and \ref{Diagrams_L}).\\

\begin{figure}
   \begin{minipage}{0.48\textwidth}
     \centering
     \includegraphics[width=1\linewidth]{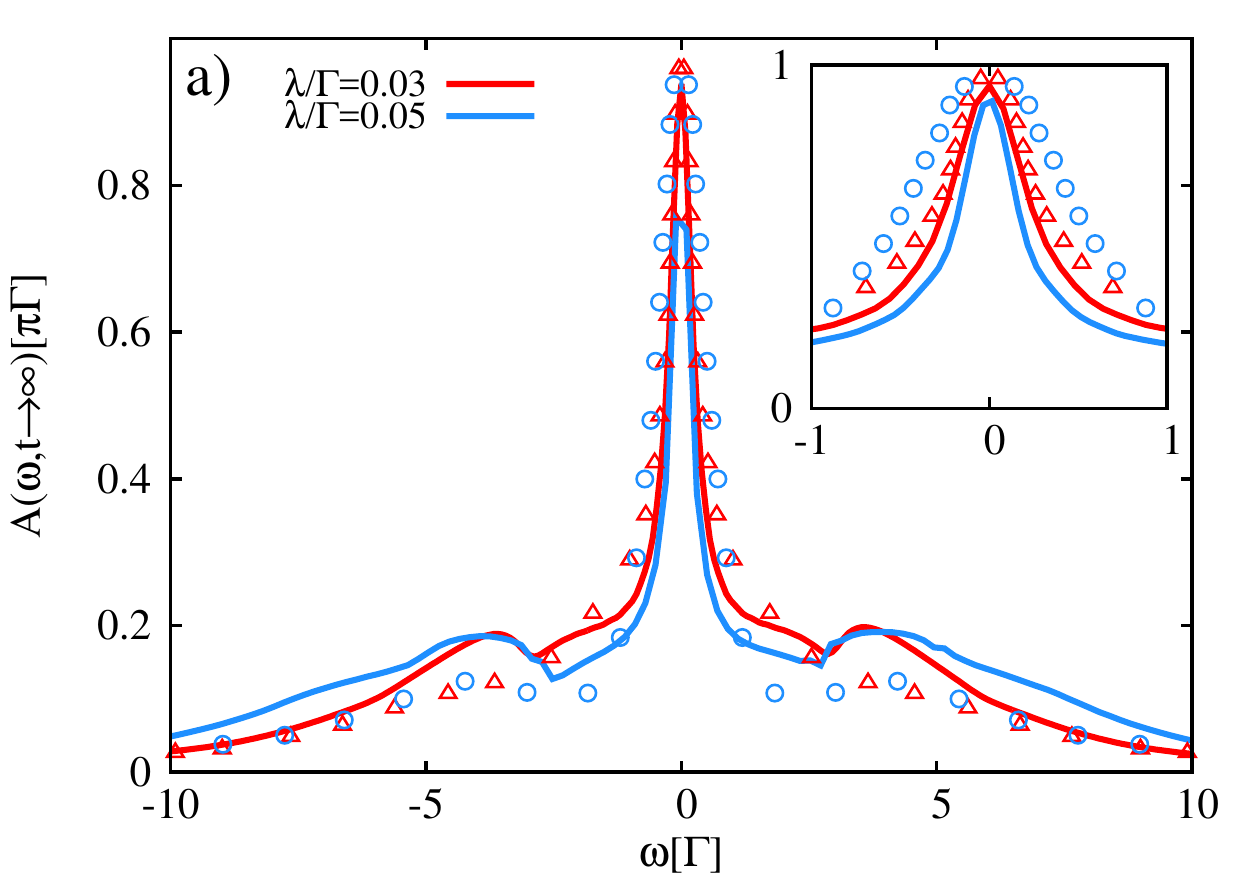}
   \end{minipage}\hfill
   \begin {minipage}{0.49\textwidth}
     \centering
     \includegraphics[width=1\linewidth]{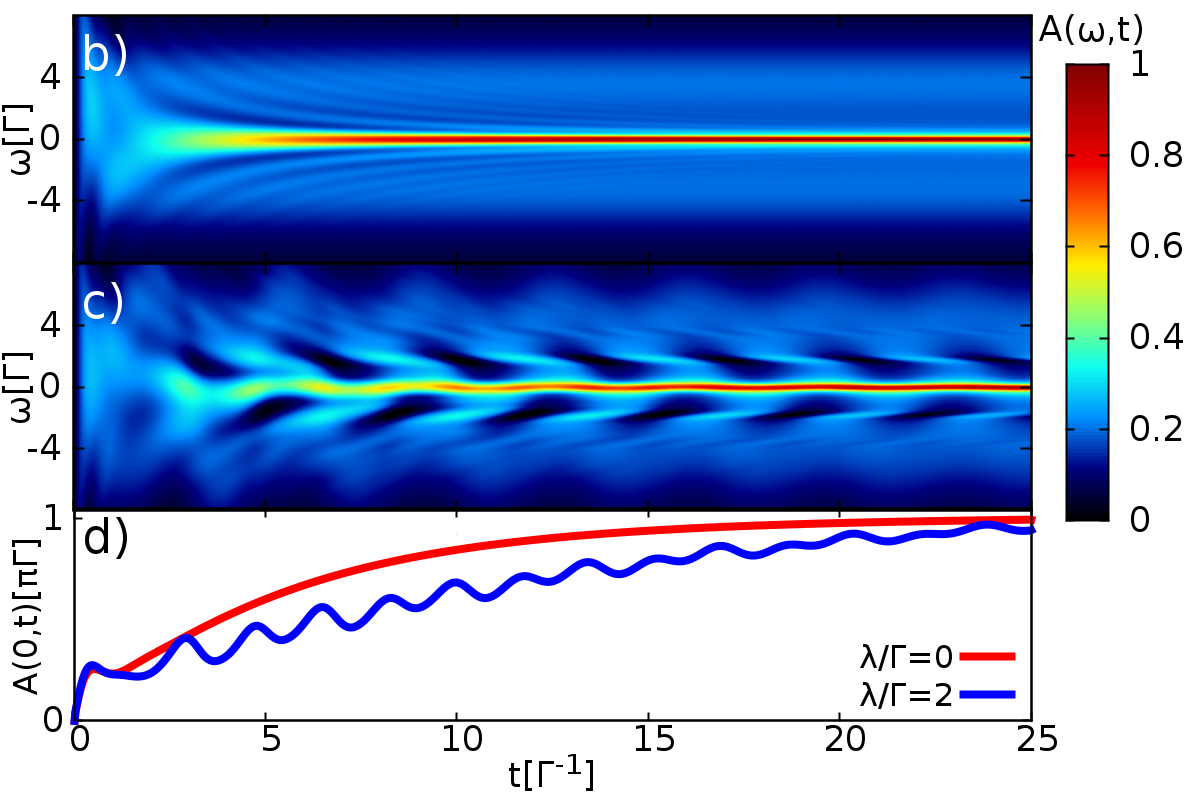}
   \end{minipage}
\caption{Spectral density in the spin-degenerate Anderson-Holstein model. a): long time values for RPA (lines) compared with equilibrium NRG results (symbols) from Ref. \cite{Jeon_PRB2003}, for two 
different electron-phonon coupling parameters: $\lambda=1.89$ (red) and $\lambda=3.14$ (blue).  The remaining 
parameters are $U=6.3$, $\omega_0=3.14$, $\epsilon_0=\lambda^2/\omega_0-U/2$ and $\Gamma_R=\Gamma_L=0.5$. b) and c): Time evolution of the density of states for $\lambda=0$ and 
$\lambda=2$. In d) we represent the central peak height evolution, showing in red the $\lambda=0$ and in blue the $\lambda=2$ case. The remaining parameters are 
$U=8$, $\omega_0=2$, $\epsilon_0=\lambda^2/\omega_0-U/2$, $\Gamma=1$ and $V=0$.}
    \label{spectral_Hewson}
\end{figure}

In Fig. \ref{spectral_Hewson} a) we show the long time spectral density compared to the exact NRG results from Ref. \cite{Jeon_PRB2003} using the RPA for $\Sigma_{e-ph}$.
Similar results are obtain for the MIGDAL approximation. As can be observed, for the smaller $\lambda$ case the RPA exhibits an overall agreement 
with the exact results. However, for 
larger values of the electron-phonon interaction the agreement becomes poorer (blue curve). In fact, this diagrammatic self-consistent approximations would not describe properly the transition to an 
insulating phase which is expected when increasing the electron-phonon interaction for $\lambda^2/\omega_0 \gtrsim U/2$ \cite{Hewson_JPC2002,Jeon_PRB2003,Martin_PRB2008}. To explore this parameter regime, 
one would need to develop an approximation correctly interpolating between the perturbative regime and the strong polaronic limit.\\


Finally, in Figs. \ref{spectral_Hewson} b) and c) we show the time evolution of the spectral density for $\lambda/\Gamma=0$ in Fig. \ref{spectral_Hewson} b) and $\lambda/\Gamma=2$ 
in Fig. \ref{spectral_Hewson} c),
 with $U/\Gamma=8$ for the RPA. We show that, even in the Kondo dominated regime, the electron-phonon interaction modifies significantly the 
system dynamics, leading to longer convergence times. This is illustrated in c) where the height of the central resonance, $A(\omega=0,t)$, is represented. We 
show that, although the central resonance width in the long time regime is not significantly modified with respect to the pure Kondo case, it exhibits different dynamical properties like oscillations 
with a period $\sim2\pi/\omega_0$. Furthermore, the decay time of these oscillations is considerably longer with respect to the $U=0$ case (not shown), indicating that the electron-electron interaction 
increases phonon retardation effects.

\section{Conclusions}
\label{sec::conclusions}
We have presented an accurate and stable algorithm to calculate the transient transport properties of interacting nanojunctions. We have shown 
how different self-consistent diagrammatic approximations can be implemented within this framework, yielding accurate results for both the transient and the steady state regimes. 
The method has allowed us to address several issues of great current interest in the condensed matter community like the dynamical build up of Kondo correlations and the possible existence of bistability in the presence of strong electron-phonon interactions.\\

For the Anderson model we have analyzed the evolution
of the spectral density explicitly exhibiting the formation of the Kondo resonance. In both cases of zero and finite voltage
bias, the results are in good agreement with available numerically exact calculations. For the electron-phonon case we have implemented two different schemes for dressing the phonon propagator 
(denoted as RPA and MIGDAL), showing the importance of a good description of the phonon dynamics to obtain accurate results. 
As a technical requirement for this implementation we have derived an expression for the  inverse of the time-discretized Keldysh free phonon propagator, allowing us to go beyond previous approaches to 
the problem based on a rotating-wave like approximation. Comparison with numerically exact results shows that the RPA and the MIGDAL approximation can provide accurate results for
the transient currents up to rather strong coupling values in the low and high voltage regimes respectively. Regarding the possible bistable behavior, we have found that electron correlation effects 
beyond the mean-field approximation tend to suppress its appearance, in agreement with recent numerically exact results \cite{Klatt_PRB2015}. However, this does not imply that upon choosing a 
different initial condition for the vibron density matrix in a model including low frequency modes, one should not observe an apparent bistability, as indicated in Refs. 
\cite{Wilner_PRB2013,Wilner_PRB2014}.\\

Finally, we have analyzed 
the situation where both interactions are present showing a reasonable agreement with the available numerically exact results for moderate electron-phonon 
coupling. We have also shown that the presence of electron-phonon interactions in the Kondo dominated regime introduces additional dynamical features in the evolution of this resonance.
We notice, however, that addressing the strong polaronic limit would require the implementation, within the present framework, of non-perturbative approximations for the self-energy in the 
spirit of Refs. \cite{Maier_PRB2011,Dong_PRB2013,Souto_PRB2014}.

\section{Acknowledgements}
R.S.S., A.L.Y. and A.M.R. acknowledge financial  support  by
Spanish MINECO through Grants No. FIS2014-55486-P and FIS2017-84860-R, and
the “Mar\'ia de Maeztu” Programme for Units of Excellence in
R\&D (Grant No. MDM-2014-0377). R.A. acknowledges support of Conseil Regional de la Nouvelle Aquitaine.
\bibliographystyle{apsrev4-1}

\bibliography{bibliography.bib}{}

\appendix
\section{Inverse free boson propagator}
\label{App::d_0}
In this appendix we discuss the problem of obtaining the inverse of the free phonon propagator discretized along the Keldysh contour. This problem has already been discussed by Kamenev in Ref. \cite{kamenev_book}, where the author considers
the problem of bosonic particles occupying a single level of energy $\omega_0$
\begin{equation}
 \hat{H}_{ph}=\omega_0b^\dagger b\,,
\end{equation}
with the free phonon propagator defined as $\hat{d}_0(t,t')=-i\left\langle T_cb(t)b^\dagger(t')\right\rangle$.
The inverse propagator in this case is formally similar to the electronic one (\ref{kamenev}), finding
\begin{equation}
i {\bold{d}}^{-1}_0 = \left(\begin{array}{cccc|cccc} -1 & & & & & & & \rho(\omega_0) \\
h_- & -1 & & & & & &  \\
& h_- & -1 & & & & &  \\
& & \ddots & \ddots & & & &  \\
\hline  
&  & & 1 & -1 & & &  \\
&  & & & h_+ & -1 & &   \\
&  &  &  & & \ddots & \ddots &  \\
&  &  & & & & h_+ & -1 \end{array} \right)_{2N\times2N} \; ,
\label{kamenev_ph}
\end{equation}
with $h_\pm=1\pm i\Delta t \omega_0 $.
This expression constitutes a discretized version of the $i\partial_t-\omega_0$ operator on the time contour with an initial condition $\rho(\omega_0)=n_p(0)/[1+n_p(0)]$, which depends on the initial 
phonon population $n_p(0)$. The obtention of the inverse free phonon propagator defined as $\hat{d}(t,t')=-i\left\langle\hat{T}_c[\hat{\varphi}(t)\hat{\varphi}^\dagger(t')]\right\rangle$, with 
$\hat{\varphi}=b+b^\dagger$ becomes more demanding since it involves the discretization of the second order differential operator $\hat{H}=p^2/2+\omega^{2}_0x^2/2$ with $p=-i\partial_x$
and $x=\sqrt{1/2\omega_0}\hat{\varphi}$. 
Moreover, it can be checked that the discretized version of the free phonon propagator given in Eq. (\ref{phonon_prop}) is not invertible as it becomes singular. In this section we discuss the way to 
obtain this inverse propagator by including a regularization procedure. By definition, the system partition function is given by \cite{kamenev_book}
\begin{equation}
 Z=\frac{\mbox{Tr}\left[\mathcal{U}_c\rho\right]}{\mbox{Tr}\left[\rho\right]}\,,
\end{equation}
where $\mathcal{U}_c=\mathcal{U}^{+}(t_{2N},t_{N+1})\mathcal{U}^{-}(t_N,t_1)$ is the contour evolution operator and $\rho=e^{-H/T}$ is the initial density matrix. Expanding 
$Z$ in coordinate space and for $N=3$ we find
\begin{equation}
 \mbox{Tr}\left[\mathcal{U}_c\rho\right]=\int dx_1\hdots dx_{6}\left\langle x_{6}\left|\mathcal{U}_{-\Delta t}\right|x_5\right\rangle\left\langle x_{5}\left|\mathcal{U}_{-\Delta t}\right|x_4\right\rangle
\left\langle x_{4}\left|\mathds{1}\right|x_3\right\rangle\left\langle x_{3}\left|\mathcal{U}_{\Delta t}\right|x_2\right\rangle
\left\langle x_{2}\left|\mathcal{U}_{\Delta t}\right|x_1\right\rangle\left\langle x_{1}\left|\rho\right|x_6\right\rangle
\end{equation}
where $\mathcal{U}_{\Delta t}=e^{-iH\,\Delta t}$. It is worth noticing that the last term in the integrand correspond to the contour closing and the third one is the branch changing in the Keldysh contour
at the final time. The relevant matrix components are given by so-called Mehler kernel \cite{Erdelyi_book}
\begin{equation}
 \left\langle x\left|e^{-iH t}\right|y\right\rangle=\frac{\exp\left\{i\left[(x^2+y^2)\cos(\omega_0t)-2xy\right]/2\sin(\omega_0t)\right\}}{\sqrt{2\pi i\sin(\omega_0 t)}}\,.
\end{equation}
Discretizing the expression and considering the time step $\Delta t$ as the smallest timescale we find
\begin{equation}
 \left\langle x\left|e^{\mp iH\Delta t}\right|y\right\rangle=\frac{\exp\left\{\pm i\left[(x^2+y^2)(1-\delta^2/2)-2xy\right]/2\delta\right\}}
{\sqrt{2\pi i\delta}}\,
\end{equation}
with $\delta=\omega_0\Delta t$. A similar expression can be found for the contour closing term
\begin{equation}
 \left\langle x\left|\rho\right|y\right\rangle=\sqrt{\frac{\rho_0}{\pi(1-\rho_{0}^2)}}\exp\left[-\frac{(1+\rho_{0}^2)(x^2+y^2)}{2(1-\rho_{0}^2)}+\frac{2xy\rho_{0}}{1-\rho_{0}^2}\right]\,,
\end{equation}
where $\rho_0=n_p(0)/[n_p(0)+1]$ contains information about the initial phonon population, $n_p(0)$. The final step for obtaining the inverse is to regularize the delta function, i.e. 
we should take
\begin{equation}
 \left\langle x\left|\mathds{1}\right|y\right\rangle\approx\frac{1}{\sqrt{2\pi\eta}}e^{-(x-y)^2/2\eta}\,,
\end{equation}
being $\eta$ an infinitesimum. Finally, the inverse of the free phonon propagator can be obtained identifying the components of
\begin{equation}
 Z=\int dx_1\hdots dx_{2N}\,e^{ix^Td^{-1}x}\,,
\end{equation}
finding the expression of Eq. (\ref{kamenev_phon}). It is worth commenting that all the prefactors in the Mehler kernel expression normalize the partition function, 
without affecting the phonon dynamics. The particular case for $N=2$ can be written as
\begin{equation}
i {\bold{d}}^{-1}_{N=2} = \left(\begin{array}{cccc} 1-\frac{\delta^2}{2}+i\delta\frac{1+\rho_{0}^2}{1-\rho_{0}^2} & -1 & 0 & -2i\frac{\delta\rho_0}{1-\rho_{0}^2} \\
-1 & 1-\frac{\delta^2}{2}+2i\frac{\delta}{\eta} & -2i\frac{\delta}{\eta} & 0  \\
0 & -2i\frac{\delta}{\eta} &  -1+\frac{\delta^2}{2}+2i\frac{\delta}{\eta} & 1\\
 -2i\frac{\delta\rho_0}{1-\rho_{0}^2} & 0 & 1 & -1+\frac{\delta^2}{2}+i\delta\frac{1+\rho_{0}^2}{1-\rho_{0}^2}
\end{array} \right) \; .
\label{kamenev_ph_N2}
\end{equation}


\end{document}